\documentclass[11pt]{article}
\usepackage{amssymb, amsmath}
\begin{document}

\makeatletter
\@addtoreset{equation}{section}
\def\theequation{\thesection.\arabic{equation}}
\def\@maketitle{\newpage
 \null
 {\normalsize \tt \begin{flushright} 
  \begin{tabular}[t]{l} \@date  
  \end{tabular}
 \end{flushright}}
 \begin{center} 
 \vskip 2em
 {\LARGE \@title \par} \vskip 1.5em {\large \lineskip .5em \begin{tabular}[t]{c}\@author 
 \end{tabular}\par} 
 \end{center}
 \par
 \vskip 1.5em} 
\makeatother
\topmargin=-1cm
\oddsidemargin=1.5cm
\evensidemargin=-.0cm
\textwidth=15.5cm
\textheight=22cm
\setlength{\baselineskip}{16pt}
\title{AdS/CFT for  3D Higher-Spin Gravity Coupled to Matter Fields}
\author{
Ippei~{\sc Fujisawa}\thanks{ifujisawa@particle.sci.hokudai.ac.jp}, Kenta~{\sc Nakagawa}\thanks{knakagawa@particle.sci.hokudai.ac.jp} and
Ryuichi~{\sc Nakayama}\thanks{nakayama@particle.sci.hokudai.ac.jp}
       \\[1cm]
{\small
    Division of Physics, Graduate School of Science,} \\
{\small
           Hokkaido University, Sapporo 060-0810, Japan}
}
\date{
EPHOU-13-010  \\
November  2013 
}
%
%
\maketitle

\begin{abstract} 
New holographic prescription for the model of 3d higher-spin gravity coupled to real matter fields  $B_{\mu\nu}$ and $C$, which was introduced in ArXiv:1304.7941[hep-th], is formulated. By using a local symmetry, two of the components of $B_{\mu\nu}$ are eliminated, and  gauge-fixing conditions are imposed such that the non-vanishing component, $B_{\phi\rho}$, satisfies a covariantly-constancy condition in the background of Chern-Simons gauge fields $A_{\mu}$, $\bar{A}_{\mu}$. In this model,  solutions to the classical equations of motion for  $A_{\mu}$ and $\bar{A}_{\mu}$ are non-flat due to the interactions with matter fields.  The solutions for the gauge fields can, however, be split into two parts, flat gauge fields ${\cal A}_{\mu}$, $\bar{{\cal A}}_{\mu}$, and those terms that depend on the matter fields. The equations for the matter fields then coincide with  covariantly-constancy equations in the flat backgrounds ${\cal A}_{\mu}$ and $\bar{{\cal A}}_{\mu}$, which are exactly the same as those in linearized 3d Vasiliev gravity.  
The two- and three-point correlation functions of operators in the boundary CFT are computed by using an on-shell action, $ \text{tr}\,  (B_{\phi\rho} \, C)$. This term does not depend on coordinates due to the matter equations of motion, and  it is not necessary to take the near-boundary limit $\rho \rightarrow \infty$. Analysis is presented for SL(3,R) $\times$ SL(3,R) as well as $HS[\frac{1}{2}] \times HS[\frac{1}{2}]$ higher-spin gravity. In the latter model, scalar operators with scaling dimensions $\Delta_+=\frac{3}{2}$ and $\Delta_-=\frac{1}{2}$ appear in a single quantization. 
\end{abstract}
\newpage
\setlength{\baselineskip}{18pt}

\newcommand{\bm}[1]{\mbox{\boldmath $#1$}}
\section{Introduction}
\hspace{5mm}
Recently, higher-spin gravity theory has been studied extensively. 
 Vasiliev et al proposed non-linear equations of motion for infinite 
tower of higher-spin gauge fields.\cite{FV1}\cite{FV2}\cite{Vasiliev}\cite{Blencowe} 
Although its description based on an action principle is still under investigation, it 
was conjectured that the higher-spin gravity in 3 dimensions is dual to the 2D W-minimal conformal field theory (CFT) models,\cite{GG}\cite{GaHa}\cite{GGHR}\cite{Ahn}\cite{GGreview} 
and this duality has been studied in the version of the model with linearized scalar 
fields.\cite{CY}\cite{KP2}\cite{AKP2}\cite{HKP}\cite{CAhn} 
Asymptotic symmetry algebra of 3d higher-spin gravity is discussed in \cite{HR}, \cite{Campoleoni}. 

It was also noticed that in 3 dimensions great simplifications occur.
The higher-spin fields can be truncated to only those with spin $s \leq N$ and the theory with 
negative cosmological constant in the frame-like approach can be defined in terms of the 
$SL(N,R) \times SL(N,R)$ Chern-Simons (CS) action.\cite{Campoleoni}  Various black hole solutions were found 
and their properties were studied.\cite{GK}\cite{AGKP}\cite{BCT}\cite{KP}\cite{KP2}\cite{GHJ}\cite{review} \cite{CLW}\cite{CLW2} 
Action integral for massless higher-spin fields in the metric-like approach was 
proposed by Fronsdal.\cite{Fronsdal} Correlation functions on the boundary conformal field theory (CFT) are studied by using 
holographic renormalization.\cite{LS} Cubic interaction vertices were also constructed.\cite{FT}  
Analysis of 3d spin-3 gravity in the metric-like approach was studied in \cite{Campometric}\cite{FN}\cite{FN2}.

Although 3D higher-spin gauge theory can be formulated in terms of the Chern-Simons (CS) theory, 
this is just a \lq pure spin-3 gravity' theory. It is desirable to include matter fields.  Actually, 
there are scalar fields in Vasiliev gravity.\cite{Vasiliev}\cite{Vasiliev3d} In \cite{FN2} two of the 
present authors proposed an action integral for matter fields interacting with higher-spin gravity. 
Matter fields are  a 0-form $C$ and a 2-form $B=\frac{1}{2} \, B_{\mu\nu} \, dx^{\mu} \wedge dx^{\nu}$. 
Hamiltonian analysis of this model was performed and it was shown that for `fixed flat' CS 
gauge fields, $A, \, \bar{A}=\omega \pm \frac{1}{\ell} \, e$, this model provides Lagrangian 
formulation of the scalars in linearlized 3d Vasiliev theory. 

In this paper we will extend the analysis of this model by treating the CS gauge fields as dynamical variables. In this model, the equations for motion of the CS gauge fields, $A_{\mu}$ and $\bar{A}_{\mu}$, are given by $F=-\frac{8\pi}{k} \, CB -(\text{trace term})$ and $\bar{F}=-\frac{8\pi}{k} \, BC  -(\text{trace term})$, where $F$ and $\bar{F}$ are field strengths, and the gauge fields are generally not flat even on shell.  On the other hand, Vasiliev gravity is a non-Lagrangian theory, and defined by classical equations of motion; flatness conditions $F=\bar{F}=0$ for gauge fields, and covariantly constancy conditions for scalars,  $dC +AC-C\bar{A}=0$ and $d\tilde{C}+\bar{A}\tilde{C}-\tilde{C}A=0$.\cite{AKP2} It has been difficult to have an explicit realization of  an action integral which yields these equations of motion. We will analyze the solutions to the equations of motion for our model, and find that they are also  solutions to the equations of motion for linearized 3d Vasiliev higher-spin gravity with flat gauge fields ${\cal A}_{\mu}$ and $\bar{{\cal A}}_{\mu}$, when the gauge fields $A_{\mu}$ and $\bar{A}_{\mu}$ are split into flat gauge fields ${\cal A}_{\mu}$ and $\bar{{\cal A}}_{\mu}$, and the parts $A_{\mu}^{BC}$ and $\bar{A}^{BC}_{\mu}$ which are written in terms of the matter fields. Hence our model may be regarded as an effective Lagrangian formulation of linearized 3d Vasiliev gravity.\footnote{See the last paragraph of sec.3.} Key point is a local symmetry of our model which allows us to eliminate two of the three components of the matter field $B_{\mu\nu}$, and to impose gauge fixing conditions which make the equations of motion for the surviving component $B_{\phi\rho}$ take the same form as those for $\tilde{C}$ in Vasiliev gravity. This is the first main result of this paper. 

Now that we have a Lagrangian formulation, we will substitute the solutions to the equations of motion into the action integral, and study AdS/CFT correspondence. The correlation functions of the boundary CFT, dual to 3d higher-spin gravity coupled with master fields, have been calculated by means of bulk-boundary  propagators.\cite{GY}\cite{CY}\cite{KP2} 
While this method is very elegant, this does not capture peculiar properties that stem from the covariant-constancy condition for matter fields, which will be discussed in the following paragraphs.  AdS/CFT is a duality between bulk AdS gravity (string) theory and CFT (large N gauge theory) on the boundary.\cite{ads1}\cite{ads2}\cite{ads3}  Explicitly, it states that the generating functional of the correlation functions in the boundary CFT is given by the on-shell action of the bulk gravity, with the boundary conditions for the bulk fields on the boundary as the source functions for the operators. Since there has been no Lagrangian formulation for 3d higher-spin gravity with matter coupling, this on-shell action method could not be employed, and  bulk-boundary propagator was used to compute two-point functions. This latter method is based on the fact that in the AdS$_3$ background the trace of the scalar field $C$ satisfies massive Klein-Gordon equation.\cite{Vasiliev}\cite{AKP2} Then, in analogy with the AdS/CFT correspondence for scalar fields in AdS space, two-point functions and the spectrum of conformal weights can be derived from the solutions with delta function boundary conditions for the scalar fields, by assuming the existence of a conjugate operator.  It would, however, be more interesting, if one could derive the same results by means of the conventional on-shell action method. Furthermore, it would be difficult, if not impossible, to generalize the method by means of the bulk-boundary propagator to fields other than scalar fields, such as spin-$\frac{1}{2}$ fields. One of the purposes of this paper is to cope with this problem. 

We found that the on-shell value of the action integral does not yield a correct generating functional of correlation functions for boundary CFT, which respect W algebra symmetry, and at the same time, that this on-shell action can be eliminated by local boundary counterterms.  We then found a new local term, $\text{tr} \, ( B_{\phi\rho}(z,\bar{z},\rho) \,  \, C(z,\bar{z},\rho))$, which yields appropriate correlation functions in the boundary CFT. Due to the equations of motion (covariantly-constancy conditions), this term turns out  independent of all coordinates $z$, $\bar{z}$, $\rho$. All data on the matter fields are stored in the internal space located at a single point in the spacetime. In some sense, holographic screen for the matter fields in this model is inside the internal space at one point in space time. Hence the near-boundary limit $\rho \rightarrow \infty$ is not necessary.  This is the second main result.  

In our BC model, there are many solutions to the classical equations of motion.  Some of them are dual to the primary operators in the boundary CFT, while the others correspond to the descendants of the primaries. These can be distinguished by calculating the three-point correlation functions, $<{\it O}(z_1) \, \bar{{\it O}}(z_2) \, J^{(s)}(z_3)>$, where ${\it O}(z_1)$ and ${\bar{\it O}}(z_2)$ are operators dual to solutions of the equations of motion, and $J^{(s)}(z_3)$, the spin-s current.  When ${\it O}(z_1)$ and ${\bar{\it O}}(z_2)$ are primary operators of W algebra, the three-point functions must satisfy appropriate relations with the two-point functions $<{\it O}(z_1) \, \bar{{\it O}}(z_2)>$. The correlation functions of scalar operators obtained agree with those obtained by means of the bulk-boundary-propagator. We also computed three-point functions of two-primary operators and a spin-s current by calculating the gauge variation of the on-shell action. Although the action integral is invariant under residual gauge transformation after gauge fixing, the solutions to the equations of motion are not, and the on-shell action changes.  

In the usual AdS/CFT correspondence for scalar fields, there exist two prescriptions for obtaining correlation functions of CFT; standard and alternate quantizations. \cite{BF}\cite{ads4} For some range of the scalar mass, there are two operators ${\it O}_{\pm}$ with scaling dimensions $\Delta_{\pm}$. Only a single operator can be realized in CFT in each quantization. In the $HS[\lambda] \times HS[\lambda]$ higher-spin gravity theory coupled to matter fields, there also exist operators ${\it O}^{(0,0)}$, ${\it O}^{(1,1)}$ and their conjugates  $\bar{{\it O}}^{(0,0)}$, $\bar{{\it O}}^{(1,1)}$ with two different  scaling dimensions in the boundary CFT. It is shown  that in our prescription for computing correlation functions with the boundary action $\text{tr} \, ( B_{\phi\rho} \,  \, C)$, both types of operators can be quantized at the same time, and the off-diagonal correlators such as  $\langle {\it O}^{(0,0)}(z_1,\bar{z}_1) \, \bar{{\it O}}^{(1,1)}(z_2,\bar{z}_2) \rangle$ vanish. So our AdS/CFT correspondence is different from the usual one for scalar fields considered in \cite{ads3}\cite{ads4}. This is the third main result. 

We also examined $HS[\frac{1}{2}] \times HS[\frac{1}{2}]$ higher-spin gravity with fermionic matter fields, by extending the matter fields to odd polynomials of the auxiliary twister variables $y_{\alpha}$. Two primary operators with spin $\frac{1}{2}$ are found, and their two-point functions and three-point functions including spin-s current are obtained by our on-shell action method. 

This paper is organized as follows. In sec.2 we will review our model for 3d higher-spin gravity coupled to 0-form ($C$) and two-form ($B$) matter fields. Its symmetry and gauge fixing procedure is explained. In sec.3 the equations of motion in the gauge-fixed form are solved. It is shown that the solution obtained here is also a solution to the equations of motion in linearized 3d Vasiliev gravity. In sec.4 the action integral including the surface term is evaluated for this solution. This on-shell action has a form which can be eliminated by appropriate local boundary counterterm. In sec.5 we choose the flat gauge fields ${\cal A}$, $\bar{{\cal A}}$ to be AdS$_3$, and evaluate the on-shell action of sec.4 for various solutions in SL(3,R) $\times$ SL(3,R) higher-spin gravity. This on-shell action does not yield appropriate two- and three-point functions. In sec.6 we propose a new boundary action, and show that this gives correct two- and three-point functions. In sec.7 our method is applied to   $HS[\frac{1}{2}] \times HS[\frac{1}{2}]$ higher-spin gravity. For this purpose it is necessary to introduce new eigenfunctions of $L_0=V^2_0$ with positive eigenvalues. In addition to two known scalar operators, ${\it O}^{(0,0)}$, ${\it O}^{(1,1)}$, which have scaling dimensions $\Delta=\frac{1}{2}$, $\frac{3}{2}$, with their conjugates $\bar{{\it O}}^{(0,0)}$, $\bar{{\it O}}^{(1,1)}$, two spin $\frac{1}{2}$ operators,  ${\it O}^{(0,1)}$, ${\it O}^{(1,0)}$ with their conjugates are found. The two- and three-point functions of these operators are computed by using the on-shell action. Summary and discussion are given in sec.8 and three appendices follow. Some formulae for sl(3,R) are collected in appendix A, new formulae for Moyal products are given in appendix B. In appendix C, scalar two-point functions in BTZ black hole are presented. 

\section{Model}
\hspace{5mm}
In this section we will review the model of 3d higher spin gravity coupled to matter fields, which was  introduced in \cite{FN2}. This model is defined by an action integral $S_{\text{CS-BC}}=S_{\text{CS}}+S_{BC}$, where $S_{\text{CS}}$ is Chern-Simons (CS) action\cite{Campoleoni}
\begin{eqnarray}
S_{\text{CS}} = \frac{k}{16\pi} \, \int_{{\cal M}} \text{tr} \, A \wedge \Big(dA+\frac{2}{3} \, A \wedge A 
\Big)-\frac{k}{16\pi} \, \int_{{\cal M}} \text{tr} \, \bar{A} \wedge \Big(d\bar{A}+\frac{2}{3} \, \bar{A} \wedge \bar{A} \Big).
\label{SCS}
\end{eqnarray} 
Here ${\cal M}=\mathbb{R} \times \Sigma$ is a 3d manifold and $\Sigma$ is a 2d manifold with boundary $\partial \, \Sigma \cong S^1$. 
In this section the properties of this model is explained by using spin-3 gravity as an example. Later in sec.7, the model extended to higher spin gravity based on $hs[\frac{1}{2}] \oplus hs[\frac{1}{2}]$ algebra will be considered. 
$A=A_{\mu} \, dx^{\mu}$ and $\bar{A}=\bar{A}_{\mu} \, dx^{\mu}$ are two SL(3,R) gauge fields.
They take values in sl(3,R) Lie algebra and can be expanded  into a basis of sl(3,R). \footnote{Our notation for sl(3,R) algebra is collected in Appendix A. } These are related to vielbein $e=e_{\mu} \, dx^{\mu}$ and spin connection $\omega=\omega_{\mu} \, dx^{\mu}$ as $A, \bar{A}=\omega \pm \frac{1}{\ell} \, e$. $\ell$ is a constant related to the cosmological constant $-\frac{2}{\ell^2}$.  The level $k$ in front of the two terms is related to the 3d Newton constant $G$ as $k=\ell/4G$. 
The second term of the action is given by 
\begin{eqnarray}
S_{BC}= \int_{{\cal M}} \text{tr} \, B \wedge \Big( dC+AC-C\bar{A}\Big). \label{SBC}
\end{eqnarray}
This is the matter action and $C$ is a real zero-form and $B=\frac{1}{2} \, B_{\mu\nu} \, dx^{\mu} \wedge dx^{\nu}$ a real two-form field. Both fields are 3 $\times $ 3 matrices and have trace components. 

We now discuss  the symmetry of the action. There are three kinds of symmetry. 
Firstly, because this action is first-order in derivatives, and is constructed as the integral of products of differential forms without an explicit 
metric, it is invariant under diffeomorphism. Secondly, it is invariant under SL(3,R) $\times$ SL(3,R) gauge transformation, with $U$ and $\bar{U}$ being SL(3,R) matrices, 
\begin{eqnarray}
A & \rightarrow & A'=U^{-1} \, dU+U^{-1} \, A \, U, \qquad 
\bar{A}  \rightarrow  \bar{A}'=\bar{U}^{-1} \, d\bar{U}+\bar{U}^{-1} \, \bar{A} \, \bar{U},  \nonumber \\
C & \rightarrow & C'=U^{-1} \, C \, \bar{U}, \qquad B \rightarrow B'=
\bar{U}^{-1} \, B \, U. \label{SL3Rmatter}
\end{eqnarray}
The diagonal SL(3,R) is local Lorentz transformation and the off-diagonal one is  local translation.\cite{Campoleoni}\cite{AT}\cite{WittenCS}.
In the pure spin-3 gravity case, the local translation coincides with the ordinary diffeomorphism and the spin-3 transformation in the metric-like formalism. When the matter fields are coupled to 
spin-3 gravity, a subgroup of the local translation in the metric-like formalism does not coincide with diffeomorphism.\cite{FN2} 

In addition to these symmetries, there is a third one. Action $S_{\text{CS-BC}}$ has the following local symmetry. 
\begin{eqnarray}
\delta_{\Xi} \, A & = & -\frac{8\pi}{k} \, \left(C \, \Xi-\frac{1}{3} \, \text{tr} \, (C\, \Xi)\right),  \nonumber\\
\delta_{\Xi} \, \bar{A} &=& -\frac{8\pi}{k} \, \left(\Xi \, C-\frac{1}{3} \, \text{tr} \, (C\, \Xi)\right), \nonumber \\
\delta_{\Xi} \, B &=& d \, \Xi+\Xi \wedge A+ \bar{A} \wedge \Xi, \nonumber \\
\delta_{\Xi} \, C &=& 0.  \label{extralocalsymmetry}
\end{eqnarray}
Here $\Xi=\Xi_{\mu} \, dx^{\mu}$ is a one-form gauge parameter function which is also a $3 \times 3$ matrix.
The trace terms on the righthand sides of $\delta_{\Xi} \, A$ and $\delta_{\Xi} \, \bar{A}$ are introduced to ensure the tracelessness of $A$ and $\bar{A}$.   The transformation (\ref{extralocalsymmetry}) with $\Xi=d\, \Sigma+\bar{A} \, \Sigma-\Sigma \, A $, where $\Sigma$ is a zero-form,  reduces on shell to SL(3,R) $\times $ SL(3,R) gauge transformation (\ref{SL3Rmatter}) with gauge parameters depending on $C\Sigma$ and $\Sigma C$.\cite{FN2}

Let us review the Hamiltonian analysis in \cite{FN2}. 
Since the 
action is first-order in derivatives, and is constructed as the integral of products of forms without an explicit 
metric, it is already in a form of Hamiltonian. By singling out the time-components of fields, we rewrite $S_{\text{CS-BC}}$ as 
\begin{eqnarray}
S_{\text{CS-BC}} &=& \int_{{\cal M}}d^3x \, \text{tr} \, \Big( -\frac{k}{16\pi} \, \epsilon^{ij} \, A_i \, \partial_t \, A_j+\frac{k}{16\pi} \, \epsilon^{ij} \, \bar{A}_i \,\partial_t \, \bar{A}_j  +\frac{1}{2} \, \epsilon^{ij} \, B_{ij} \, \partial_t \, C        \Big) \nonumber \\
&& +\frac{k}{16\pi} \, \int_{{\cal M}} d^3x \, \text{tr} \, \Big(A_t \, \psi-\bar{A}_t \, \bar{\psi} \Big)+\int_{{\cal M}} d^3x \, \epsilon^{ij} \, \text{tr} \, B_{ti} \, \chi_j \nonumber \\
&&+\frac{k}{16\pi} \, \int_{\partial {\cal M}} \, dt \, d\phi \, \text{tr} \, \Big(A_t \, A_{\phi}-\bar{A}_t \, \bar{A}_{\phi} \Big). \label{SHamiltonian}
\end{eqnarray}
Here $\epsilon^{ij}$ is a Levi-Civita symbol for the spatial components, and $\epsilon^{\phi \rho}=+1$. 
The coordinates are $t, \phi$ and $\rho$. The last term is a boundary term on  $\partial {\cal M}=\mathbb{R}_t \times S^1_{\phi}$, which appeared after partial integration. Functions $\psi$, $\bar{\psi}$ and $\chi_i$ are defined as 
\begin{eqnarray}
\psi & =& \epsilon^{ij} \, (F_{ij}+\frac{8\pi}{k} \, C \, B_{ij}-\frac{8\pi}{3k} \, \text{tr}(CB_{ij})) , 
\label{psi} \\
\bar{\psi} &=& \epsilon^{ij} \, (\bar{F}_{ij}+\frac{8\pi}{k} \, B_{ij} \, C-\frac{8\pi}{3k} \, \text{tr}(CB_{ij})), \label{psibar} \\
\chi_i &=& \partial_i \, C+A_i \, C-C \, \bar{A}_i. \label{chii}
\end{eqnarray}
Here $F=dA+A \wedge A$ and $\bar{F}=d\bar{A}+\bar{A} \wedge \bar{A}$ are field strengths.

The momentum $P$ conjugate to $C$ is given by $P\equiv  \frac{1}{2} \, \epsilon^{ ij} \, B_{ij}$. 
The momenta conjugate to 
$A_i$ and $\bar{A}_i$ are  $\pi_A^i \equiv \frac{k}{16\pi} \, \epsilon^{ij} \, A_j$ and $\pi_{\bar{A}}^i \equiv \frac{-k}{16\pi} \, 
\epsilon^{ij} \, 
\bar{A}_j$, respectively. The momentum $\Pi_{B}^i$ conjugate to $B_{ti}$ does not exist, and the primary constraint 
$\Pi_{B}^i \approx 0$ generates  a secondary constraint $\chi_i \approx 0$. 
Similarly, the momenta  $\pi_A$ and $\pi_{\bar{A}}$ conjugate to $A_t$ and $\bar{A}_t$, respectively, obey $\pi_{A} 
\approx \pi_{\bar{A}} \approx 0$.  These lead to secondary constraints, $\psi \approx 0$ and $\bar{\psi} \approx 0$. 
The Hamiltonian is a sum of Lagrange multipliers times these constraint functions. 

Constraints $\pi_A \approx \pi_{\bar{A}} \approx \Pi_B^i \approx 0$ are first-class. This means that $A_t$, $\bar{A}_t$, $B_{ti}$, as well as $\pi_A$, $\pi_{\bar{A}}$, $\Pi_B^i$, are unphysical. So, by means of  (\ref{extralocalsymmetry}), we can gauge fix  $B_{t\phi}$, $B_{t\rho} $ such that  $B_{t\phi}=B_{t\rho}=0$. The two constraints $\psi$ and $\bar{\psi}$ generate $SL(3,R) \times SL(3,R)$ gauge transformations, and are first-class. The  function $\chi_i$, the generator of the local transformation (\ref{extralocalsymmetry}),  transforms covariantly under these gauge  transformations, hence $\chi_i \approx 0$ is also first-class. Now, the gauge fields $A_i$ are eliminated by $\psi \approx 0$ and an appropriate gauge fixing, and are non-propagating in the bulk. So are $\bar{A}_i$. As for $\chi_i$, the role of these constraints is not to eliminate $C$, but to determine the derivatives of the field $C$ in the spatial directions, $\phi$ and $\rho$. Corresponding to $\chi_i$, we thus propose to fix gauge by the conditions 
\begin{equation}
\tilde{\chi}_i \equiv \partial_i \, P-P \, A_i+\bar{A}_i \, P = 0. \label{chitilde}
\end{equation}
In this way,  fields $C$ and $P$ are determined by their values at a sinlgle point in space time. The number of constraints is larger than that of fields, and there are no physical degrees of freedom.\footnote{The analysis in this section also applies to HS[$\lambda$] $\times$ HS[$\lambda$] higher-spin gravity with $0 \leq \lambda \leq 1$. In this case, however,  the internal space becomes infinite-dimensional. This means there are an infinite number of fields, and even in the presence of the constraint $\chi_i=0$, there remains a propagating degree of freedom, $\text{tr} \, C$.  In AdS space, $\text{tr} \, C$ satisfies Klein-Gordon equation and has a non-polynomial solution with a delta-function boundary condition at space-like infinity. }  
Combining the gauge fixing $\tilde{\chi}_i \approx 0$ with the equation of motion for $P$,\footnote{Recall that we set $B_{t\phi}=B_{t\rho}=0$. The equation of motion for $B_{\mu\nu}$ is (\ref{eomB}) below.} 
we obtain the set of  equations for $P$, 
\begin{equation}
\partial_{\mu} \, P-P \, A_{\mu}+\bar{A}_{\mu} \, P=0.
\end{equation}
 This provides the counterpart of the equations of motion for $C$, $\partial_{\mu} \, C+C \, A_{\mu}-\bar{A}_{\mu} \, C=0$.

\section{Solution to the Equations of Motion}
\hspace{5mm}

The equations of motion are given by 
\begin{eqnarray}
F &=& -\frac{8\pi}{k} \, (CB-\frac{1}{3} \, \text{tr} \, CB), \label{FCB}\\
\bar{F} &=& -\frac{8\pi}{k} \, (BC-\frac{1}{3} \, \text{tr} \, CB),  \label{FbBC}
\end{eqnarray}
and
\begin{eqnarray}
d \, C+A C-C \bar{A} &=& 0, \\
d \, B+\bar{A} \wedge B-B \wedge A &=& 0.  \label{eomB}
\end{eqnarray}
As discussed in the previous section, we adopt the gauge fixing conditions
\begin{equation}
B_{t\rho}=B_{t\phi}=0, \quad \tilde{\chi}_i=\partial_i \, B_{\phi \rho}-B_{\phi \rho} \, A_i+\bar{A}_i \, B_{\phi \rho} =0. \label{BB}
\end{equation}
Then the equations of motion for the gauge fields are 
\begin{eqnarray}
F_{t\rho} &=& 0, \label{Ftrho} \\
F_{t\phi} &=&0, \label{Ftp} \\
F_{\phi\rho} &=& -\frac{8\pi}{k} \, (CB_{\phi \rho}-\frac{1}{3} \, \text{tr} \, CB_{\phi \rho}), \label{Frhophi}
\end{eqnarray}
There are similar equations for $\bar{A}$. The equations for $B_{\phi\rho}$ are given by 
\begin{eqnarray}
\partial_{\mu} \, B_{\phi\rho} +\bar{A}_{\mu} \, B_{\phi\rho}-B_{\phi\rho} \, A_{\mu}=0. \label{eomBpr}
\end{eqnarray}
These are similar to those for $C$, 
\begin{eqnarray}
\partial_{\mu} \, C +A_{\mu} \, C-C \, \bar{A}_{\mu}=0.  \label{eomC}
\end{eqnarray}

To solve these equations, we need to gauge fix $A$, $\bar{A}$. This is done by the conditions 
\begin{eqnarray}
A_{\rho} &= &b^{-1}(\rho) \, \partial_{\rho} \, b(\rho), \label{gaugeArho} \\
\bar{A}_{\rho} &=& b(\rho) \, \partial_{\rho} \, b^{-1}(\rho). \label{gaugeAbarrho}
\end{eqnarray}
Here $b(\rho)$ is a function of only $\rho$. This gauge fixing is always possible.\cite{Campoleoni} 

The equation for $A_t$, $\partial_{\rho} \, A_t=-[A_{\rho}, A_t]$, which follows from (\ref{Ftrho}), can now be solved to yield  
\begin{equation}
A_t = b^{-1}(\rho) \, a_t(t,\phi) \, b(\rho).  \label{Atbatb}
\end{equation}
$a_t$ is an arbitrary function of $t$ and $\phi$.  The equation for $A_{\phi}$, (\ref{Frhophi}), can be rewritten as 
\begin{equation}
\partial_{\rho} \, (b \, A_{\phi} \, b^{-1}) = \frac{8\pi}{k} \, \Big(b \, C \, B_{\phi\rho} \, b^{-1}-
\frac{1}{3} \, \text{tr} \, C \, B_{\phi\rho}\Big). \label{bAb}
\end{equation}
One can show that the righthand side is independent of $\rho$, because $\partial_{\rho} \, (b C \, b)=0$ due to (\ref{eomC}) and $\partial_{\rho} \, (b^{-1} B_{\phi\rho} \, b^{-1})=0$ due to 
(\ref{eomBpr}). Hence the solution to (\ref{bAb}) is obtained as
\begin{eqnarray}
A_{\phi} =b^{-1} \, a_{\phi}(t,\phi) \, b+\frac{8\pi}{k} \, (\rho-\rho_0) \, \Big(C \, B_{\phi\rho} -
\frac{1}{3} \, \text{tr} \, C \, B_{\phi\rho}\Big).  \label{Aphi}
\end{eqnarray}
Here $\rho_0$ is a constant, which parametrize the solution,  and $a_{\phi}$ is an arbitrary function of $t$ and $\phi$. 
The second term on the right corresponds to a torsion.\cite{FN2}
Finally,  (\ref{Ftp}) follows, if $a_t$ and $a_{\phi}$ satisfy an equation,
\begin{eqnarray}
&& \partial_t \, a_{\phi}-\partial_{\phi} \, a_t+[a_t, a_{\phi}]  \label{atp}\\
&& +\frac{8\pi (\rho-\rho_0)}{k} \, \partial_t \, (b \, C \, B_{\phi\rho} \, b^{-1}-\frac{1}{3} \, \text{tr} \, C \, B_{\phi\rho})+\frac{8\pi (\rho-\rho_0)}{k} \, [a_t, b\, C  \, B_{\phi\rho}\, b^{-1}]=0 \nonumber 
\end{eqnarray}
Now by using (\ref{eomC}) we can show that 
\begin{equation}
\partial_t \, (bCb) +a_t \, (bCb)-(bCb) \, \bar{a}_t=0.  \label{tbCb}
\end{equation}
Here $\bar{a}_t$ is defined later in (\ref{Atbarbabarb}) by solving for $\bar{A}_t$ as in (\ref{Atbatb}). 
Combining this with a similar equation for $B_{\phi\rho}$,  
\begin{equation}
\partial_t \, (b^{-1} \, B_{\phi\rho} \, b^{-1}) +\bar{a}_t \, (b^{-1} \, B_{\phi\rho} \, b^{-1})-(b^{-1} \, B_{\phi\rho} \, b^{-1}) \, a_t=0,
\end{equation}
 which is obtained from (\ref{eomBpr}), we find that 
\begin{equation}
\partial_t \, (bCB_{\phi\rho}b^{-1})+a_t \, (bCB_{\phi\rho}b^{-1})-(bCB_{\phi\rho}b^{-1}) \, a_t=0.
\end{equation}
Trace of this equation leads to $\partial_t \, \text{tr} \, (CB_{\phi\rho})=0$. Hence the terms in the second line of (\ref{atp}) cancel out: 
\begin{equation}
\partial_t \, a_{\phi}-\partial_{\phi} \, a_t+[a_t, a_{\phi}] =0. \label{flattphi}
\end{equation}
Therefore except for the extra term  (\ref{Aphi}) in $A_{\phi}$, the solution for the gauge field is the same as that in the pure spin-3 gravity obtained in \cite{Campoleoni}.  In the following, we will set a gauge condition
\begin{equation}
a_-(t,\phi) \equiv \frac{1}{2} \, (a_t(t,\phi)-a_{\phi}(t,\phi))=0. \label{gaugea}
\end{equation}
Equation (\ref{flattphi}) is then satisfied, if $a_{\phi}$ is holomorphic, $a_{\phi}=a_{\phi}(x^+)$:\footnote{$x^{\pm} \equiv t \pm \phi$.}
\begin{equation}
\partial_- \, a_{\phi}  \equiv \frac{1}{2} \, (\partial_t-\partial_{\phi}) \,  a_{\phi}=0. 
\label{del-aphi}
\end{equation} 

Here we pause for a moment to consider the boundary conditions on the gauge fields. To make the variation problem well-defined, we must impose 
appropriate boundary conditions. 
When we vary the action with respect to $A$, $\bar{A}$, $B$ and $C$, we obtain boundary terms
\begin{eqnarray}
\delta \, S_{\text{CS-BC, boundary}} = -\frac{k}{16\pi} \, \int_{\partial {\cal M}} \, \text{tr} \, (A \wedge \delta \, A-\bar{A} \, \wedge \delta \, \bar{A}).
\end{eqnarray}
The variation term $\int_{\partial {\cal M}} \, \text{tr} \, B \, \delta \, C$ drops, because $B_{t\phi}=0$. 
In the pure spin-3 gravity, natural boundary conditions were \cite{Campoleoni} 
\begin{equation}
A_-\big|_{\text{boundary}}= \bar{A}_+\big|_{\text{boundary}}=0. \label{Aminus}
\end{equation}
Then the gauge fixing (\ref{gaugea}) ensures $A_-=0$ everywhere on shell.\cite{Campoleoni}
When the matter fields are coupled, however, we obtain from (\ref{Atbatb}) and (\ref{Aphi}) that 
\begin{equation}
A_-=-\frac{4\pi}{k} \, (\rho-\rho_0) \, \Big( C \, B_{\phi\rho}-\frac{1}{3} \, \text{tr} \, C\, B_{\phi\rho} \Big). \label{Amin}
\end{equation}
This does not vanish as $\rho \rightarrow \infty$, and we cannot adopt the boundary conditions (\ref{Aminus}). 
Instead, we impose the following conditions.
\begin{eqnarray}
& A_-\big|_{\text{boundary}} \rightarrow -\frac{4\pi}{k} \, (\rho-\rho_0) \, \Big( C \, B_{\phi\rho}-\frac{1}{3} \, \text{tr} \, C\, B_{\phi\rho} \Big),    \label{Aminustrue} \\
& (C, \ B_{\phi\rho})|_{\text{boundary}}=\text{fixed} \qquad (\rho \rightarrow \infty) \label{Aminustrue2}
\end{eqnarray}
As discussed at the end of sec.2, fields $C$ and $B_{\phi \rho}$ on the boundary are completely determined in terms of their values at a single point in the bulk by the covariant-constancy conditions (\ref{eomBpr}) and (\ref{eomC}). 
Hence the boundary condition (\ref{Aminustrue2}) will be appropriate.  When the vielbein is computed from $A$ and $\bar{A}$, the spacetime is not asymptotically AdS. Furthermore, we need to introduce surface terms on the time-like boundary.  
\begin{equation}
S_{\text{surface}} = \frac{k}{8\pi} \, \int_{\partial {\cal M}} \, dt \, d\phi \, \text{tr} \, \big(A_+ \, A_-+\bar{A}_+ \, \bar{A}_-\big)  \label{surface}
\end{equation}
Then the  variation of $S_{\text{CS}}+S_{\text{surface}}$ is given by 
$\frac{k}{4\pi} \, \int_{\partial {\cal M}} \, dt \, d\phi \, \text{tr} \, \big(A_+ \, \delta \, A_-  +\bar{A}_- \, \delta \, \bar{A}_+\big) $. 
This vanishes, when the gauge field and  matter fields are fixed as in  (\ref{Aminustrue}) and (\ref{Aminustrue2}). 

Parameter of gauge transformation which preserves the boundary condition (\ref{Aminustrue}) is given by 
\begin{equation}
\Lambda = b^{-1} \, \lambda(x^+) \, b.  \label{residualsym}
\end{equation}
Since $\partial_- \, \lambda=0$, $A_-$ transforms as $\delta \, A_-= [A_-, \Lambda]$.  Because matters transform as $\delta \, C=-\Lambda \, C$ and $\delta B_{\phi\rho}= B_{\phi\rho} \, \Lambda$,  the boundary condition (\ref{Aminustrue}) is covariant under (\ref{residualsym}). 
However, the condition  (\ref{Aminustrue2}) is not invariant. 

As for $\bar{A}$, by repeating steps similar to those for $A$, we obtain (\ref{gaugeAbarrho}) and 
\begin{eqnarray}
\bar{A}_t &=& b \, \bar{a}_t \, b^{-1}, \label{Atbarbabarb}\\
\bar{A}_{\phi} &=& b \, \bar{a}_{\phi}(t,\phi) \, b^{-1}+\frac{8\pi}{k} \, (\rho-\rho_0) \, \Big( B_{\phi\rho} \, C-
\frac{1}{3} \, \text{tr} \, C \, B_{\phi\rho}\Big). \label{Aphibar}
\end{eqnarray}
Here for simplicity we introduced the same integration constant $\rho_0$ as in (\ref{Aphi}). 
$\bar{a}_i$ must satisfy relations
\begin{equation}
\bar{a}_{\phi} =-\bar{a}_t, \qquad \partial_+ \, \bar{a}_{\phi}=0. \label{gaugeabar}
\end{equation}
The boundary condition for $\bar{A}_+$ is 
\begin{equation}
\bar{A}_+\big|_{\text{boundary}} \rightarrow \frac{4\pi}{k} \, (\rho-\rho_0) \, \Big( B_{\phi\rho}\, C-\frac{1}{3} \, \text{tr} \, C\, B_{\phi\rho} \Big). \qquad (\rho \rightarrow \infty) \label{Abarplustrue}
\end{equation}

Finally, we will solve for $C$ and $B_{\phi\rho}$.   By using (\ref{eomC}) we have (\ref{tbCb}) and 
\begin{equation}
\partial_{\phi} \, (bCb)+a_{\phi} \, (bCb)-(bCb) \, \bar{a}_{\phi}=0,  \label{phibCb}
\end{equation}
where the terms of $A_{\phi}$ and $\bar{A}_{\phi}$ which depend on $C$ and $B_{\phi\rho}$ canceled out. 
Eqs (\ref{tbCb}) and (\ref{phibCb}) yield the equations
\begin{eqnarray}
\partial_- \, (b \, C \, b) &=&  - (b\, C \, b) \, \bar{a}_{\phi}, \label{del-bCb}\\
\partial_+ \, (b \, C \, b) &=& -  a_{\phi}\, (b \, C \, b). 
\end{eqnarray}
These equations are solved in terms of the ordered exponential $({\cal P} \, \exp)$ and anti-ordered exponential $(\overline{{\cal P}} \, \exp)$. 
\begin{equation}
C = b^{-1} \, \Bigg({\cal P} \, \exp \Big[ -\int^{x^+}_{x^+_1} dx^+ \, a_{\phi}(x^+) \Big] \Bigg)\, C(0) \, \Bigg(\overline{{\cal P}}\, \exp \Big[ -\int^{x^-}_{x^-_1} dx^- \, \bar{a}_{\phi}(x^-) \Big]\Bigg)  \, b^{-1} \label{Csol}
\end{equation}
Here $C(0)$ is a constant matrix, and $x_1^{\pm}$ constants.  In exactly the same manner, we obtain $\partial_- \, (b^{-1} \, B_{\phi\rho} \, b^{-1})=\bar{a}_{\phi} \, b^{-1} \, B_{\phi\rho} \, b^{-1}$ and 
$\partial_+ \, (b^{-1} \, B_{\phi\rho} \, b^{-1})=b^{-1} \, B_{\phi\rho} \, b^{-1} \, a_{\phi}$,  
which are solved for $B_{\phi\rho}$. 
\begin{equation}
B_{\phi\rho} = b \, \Bigg({\cal P} \, \exp \Big[ \int^{x^-}_{x^-_2} dx^- \, \bar{a}_{\phi}(x^-) \Big]\Bigg) \, B_{\phi\rho}(0) \, \Bigg(\overline{{\cal P}}\, \exp \Big[ \int^{x^+}_{x^+_2} dx^+ \, a_{\phi}(x^+) \Big]\Bigg)  \, b  \label{Bsol}
\end{equation}
Hence solutions  for $C$ and $B_{\phi\rho}$ are expressed in terms of  flat gauge fields ${\cal A}_{\mu} \equiv b^{-1} \, a_{\mu} \, b+b^{-1} \, \partial_{\mu} \, b$ and $\bar{{\cal A}}_{\mu} \equiv b \, \bar{a}_{\mu} \, b^{-1}+b \, \partial_{\mu} \, b^{-1}$. ($a_{\rho}=\bar{a}_{\rho}=0$)

To summarize, the solution obtained in this section also satisfies the following equations. (${\cal A} \equiv {\cal A}_{\mu} \, dx^{\mu}$)
\begin{eqnarray}
{\cal F}  \equiv  d{\cal A}+{\cal A} \wedge {\cal A} &=& 0, \\
\bar{{\cal F}}  \equiv  d \bar{{\cal A}}+\bar{{\cal A}} \wedge \bar{{\cal A}}&=&0, \\
d \, C+{\cal A} \, C-C \, \bar{{\cal A}} &=&0, \\
d \, B_{\phi\rho}+\bar{{\cal A}} \, B_{\phi\rho}-B_{\phi\rho} \, {\cal A} &=&0 \label{Bcovconst}
\end{eqnarray}
These equations resemble those of 3d Vasiliev gravity with linearlized interaction. \cite{Vasiliev3d}\cite{AKP2} 
There is a difference, however. In \cite{AKP2}, field $C$ is a complex scalar, and a conjugate of $C$ is introduced and denoted as $\bar{C}$. The holographic duals of the traces of these fields, ${\it O}$ and $\bar{{\it O}}$, are conjugate to each other, and have non-vanishing two-point function. There also exists another complex scalar $\tilde{C}$, whose equation of motion is same as  (\ref{Bcovconst}) for  $B_{\phi\rho}$. There also exists a conjugate of $\tilde{C}$, {\em i.e.}, $\stackrel{\eqsim}{C}$, and the traces of both fields have holographic duals, $\tilde{O}$ and $\stackrel{\eqsim}{\it O}$, with non-vanishing two-point function. In our model, $C^{\text{ours}}$ and $B_{\phi\rho}$ 
are real fields, and $B_{\phi\rho}$  plays the role of both $\tilde{C}$ and $\bar{C}$.  It might seem that, to make the correspondence of the two models exact, we need to double our real scalars.
It will, however, be shown in sec.7 that it is possible to obtain all necessary two-point functions in our model without introducing extra matter fields.

\section{On-shell Action}
\hspace{5mm}
In AdS/CFT correspondence, a generating functional of the correlation functions on the boundary CFT is obtained by 
substituting solutions to the equations of motion into the action integral $S_{\text{CS-BC}}+S_{\text{surface}}$.\cite{ads1}\cite{ads2}\cite{ads3} Let us compute this on-shell action. 

When the solutions to the equations of motion are substituted, $S_{\text{BC}}$ (\ref{SBC}) vanishes. As for $S_{\text{CS}}$ 
(\ref{SCS}), by rewriting $dA+ \frac{2}{3} \, A \, \wedge \, A$ as $F-\frac{1}{3} \, A \, \wedge \, A$ and using (\ref{FCB}), 
we obtain 
\begin{equation}
S_{\text{CS (on shell)}} = \frac{k}{16\pi} \, \int \text{tr} \, A \wedge \big(-\frac{8\pi}{k} \, CB-\frac{1}{3} \, A \wedge A\big)-
\frac{k}{16\pi} \, \int \text{tr} \, \bar{A} \wedge \big(-\frac{8\pi}{k} \, BC-\frac{1}{3} \, \bar{A} \wedge \bar{A}\big).  
\end{equation}
By using (\ref{BB})  this is rewritten as $-\frac{1}{2}\, \int_{ {\cal M}} \text{tr} \, (A_t \, C \, B_{\phi\rho}) \, d^3x-\frac{k}{16\pi} \, \int_{ {\cal M}} 
\text{tr} \, [A_t,A_{\phi}] \, A_{\rho} \, d^3x$+ (similar terms for $\bar{A}$). Owing to (\ref{Ftp}), the integrand of the second term becomes total derivative terms, $\text{tr} \, (\partial_{\phi} \, A_t-\partial_t \, A_{\phi}) \,  A_{\rho}(\rho)=\text{tr} \,  \big(\partial_{\phi} \, (A_t \, A_{\rho})-\partial_t \, (A_{\phi} \, A_{\rho})\big)$, hence the integral 
vanishes. We obtain 
\begin{multline}
 S_{\text{CS (on shell)}} \\
= -\frac{1}{2} \, \rho_c \, \int_{\rho=\rho_c } \, dt \, d\phi \, \text{tr} \, (A_t \, C \, B_{\phi\rho}) +\frac{1}{2} \, \rho_c \, \int_{\rho=\rho_c} \, dt \, d\phi \, \text{tr} \, (\bar{A}_t \, B_{\phi\rho} \, C).  \label{1}
\end{multline}
Here we used the fact that the trace terms do not depend on  $\rho$, and the $\rho$ integrations amount to simply multiplying the integrands by $\rho_c$, where  $\rho_c$ is a large cut-off representing the location of the boundary. 

The surface term (\ref{surface}) on shell is given by 
\begin{eqnarray}
S_{\text{surface (on shell)}} &=& -\frac{1}{2} \, (\rho_c-\rho_0) \, \int_{\rho=\rho_c} dt \, d\phi \, \text{tr} \, (a_{\phi} \, b \, C \, B_{\phi\rho} \, b^{-1}) \nonumber \\
&&-\frac{1}{2} \, (\rho_c-\rho_0) \, \int_{\rho=\rho_c} dt \, d\phi \, \text{tr} \, (\bar{a}_{\phi} \, b^{-1}  \, B_{\phi\rho} \, C \, b)  \nonumber \\
&& -\frac{4\pi}{k} \, (\rho_c-\rho_0)^2 \, \int_{\rho=\rho_c}  dt \, d\phi \, \big(\text{tr} (B_{\phi\rho} \, C)^2-\frac{1}{3} \, (\text{tr} B_{\phi\rho} \, C)^2 \big).\label{2}
\end{eqnarray}
By adding (\ref{1}) and (\ref{2}), we obtain the on-shell action
\begin{eqnarray}
&& \, \int_{\rho=\rho_c} dt  d\phi \Big( -   (\frac{1}{2} \,\rho_0-\rho_c) \ \text{tr} \, \big( B_{\phi\rho} \, \partial_+ \, C) 
  - (\frac{1}{2} \,\bar{\rho}_0-\rho_c) \text{tr} \,   \big(   B_{\phi\rho}\, \partial_-C\big)  \Big)  \nonumber \\
&&    -\frac{4\pi}{k} \, (\rho_c-\rho_0)^2 \, \int_{\rho=\rho_c}  dt \, d\phi \, \big(\text{tr} (B_{\phi\rho} \, C)^2-\frac{1}{3} \, (\text{tr} B_{\phi\rho} \, C)^2 \big)                      . 
\end{eqnarray}
Here (\ref{Csol}) and (\ref{Bsol}) are used.
This is divergent as $\rho_c \rightarrow \infty$, but the divergence can be cancelled by local boundary counterterms
\begin{eqnarray}
S_{\text{counterterms}} &=& -(\rho_c-\rho_1) \, \int_{\rho=\rho_c} dt  d\phi \ \text{tr} \, \big( B_{\phi\rho} \, \partial_+ \, C)  \nonumber \\
&&  - ( \rho_c-\bar{\rho}_1) \, \int_{\rho=\rho_c} dt \, d\phi \ \text{tr} \,\big(   B_{\phi\rho}\, \partial_-C\big)  \nonumber \\
&&   +\frac{4\pi}{k} \, (\rho_c-\rho_0)^2 \, \int_{\rho=\rho_c}  dt \, d\phi \, \big(\text{tr} (B_{\phi\rho} \, C)^2-\frac{1}{3} \, (\text{tr} B_{\phi\rho} \, C)^2 \big). \label{localcounterterm}
\end{eqnarray}
$\rho_1$ and $\bar{\rho}_1$ are some finite constants. Finally, the on-shell action which includes the counterterms becomes finite. It can be re-expressed in a simpler form. 
\begin{eqnarray}
S_{\text{on shell}} 
 &=& \mu \, \int_{\rho=\rho_c} dt  d\phi \ \text{tr} \, \big( B_{\phi\rho} \, \partial_+ \, C) 
 + \bar{\mu} \, \int_{\rho=\rho_c} dt \, d\phi \ \text{tr} \, \big(   B_{\phi\rho}\, \partial_-C\big), \label{renormalizedonshellaction}
\end{eqnarray}
where $\mu=\rho_1-\frac{1}{2} \, \rho_0$ and $\bar{\mu}=\bar{\rho}_1-\frac{1}{2} \, \rho_0$.  The quartic terms are eliminated, because the theory is free.   
By using (\ref{del-aphi}), (\ref{del-bCb}) and a similar equation for $b^{-1}B_{\phi\rho}b^{-1}$,  it can be shown that the integrand of the first term of  (\ref{renormalizedonshellaction}) is independent of $x^-$. Similarly, that of the second term does not depend on $x^+$, either. 

There remains ambiguity in the finite coefficients in front of the above two integrals.   Notice that by setting $\rho_0=2\rho_1$ and  $\rho_0=2 \, \bar{\rho}_1$, we can drop both terms. In  sec.5 we will show that this boundary action does not yield appropriate correlation functions which satisfy W-algebra Ward identities.  We will then drop this on-shell action completely by using the local counterterms, and introduce another local boundary action integral in sec.6. In this way the residual gauge symmetry (\ref{residualsym})  will be recovered.

\section{Explicit Solution for $C$ and $B$ in AdS$_3$ Backgrounds}
\hspace{5mm}
As we have seen, although the gauge fields $A$, $\bar{A}$ are not flat in the matter-coupled theory, the fields $a$, $\bar{a}$ (${\cal A}_{\mu}$, $\bar{{\cal A}}_{\mu}$) are flat, and it may make sense to discuss matter fields associated to AdS$_3$ background. As a warming up, in this section we will compute the solutions for these fields in AdS$_3$, and evaluate the on-shell action (\ref{renormalizedonshellaction}) in SL(3,R) $\times$ SL(3,R) theory. In sec.7 higher spin gravity with $HS[\frac{1}{2}] \times HS[\frac{1}{2}]$ gauge symmetry will be considered. 

For asymptotically AdS$_3$ spacetime, $a$ and $\bar{a}$ are given by \cite{Bana}
\begin{eqnarray}
a &= & (L_1+\frac{2\pi}{k} \, {\cal L}(x^+) \, L_{-1}) \, dx^+,  \\
\bar{a} &= & -(L_{-1}+\frac{2\pi}{k} \, \bar{{\cal L}}(x^-) \, L_{1}) \, dx^-,  \label{asmAdS3}
\end{eqnarray}
and $b(\rho)$ by 
\begin{equation}
b(\rho)=e^{\rho \, L_0}. 
\end{equation}
For matrices $L_i$ see appendix A. The AdS$_3$ spacetime in the Poincar\'{e} patch is given by ${\cal L}=\bar{{\cal L}}=0$,  and the one in the global patch is obtained by the choice $\frac{2\pi}{k} \, {\cal L}= \frac{2\pi}{k} \, \bar{{\cal L}} =\frac{1}{4}$.  
In what follows we will consider AdS$_3$ spacetime in the Poincar\'{e} patch. So we
set $a=L_1 \, dx^+$ and $\bar{a}=-L_{-1} \, dx^-$, and the metric is given by $ds^2=\ell^2 (d\rho^2+e^{2\rho} \, (-dt^2+d\phi^2))$. 

We Wick rotate the spacetime to Euclidean AdS$_3$, by replacements $x^+=t+\phi \rightarrow \phi-i\tau=z$, $x^-=t-\phi \rightarrow -(\phi+i\tau)=-\bar{z}$. From (\ref{Csol}) and (\ref{Bsol}) we write down the formulae for $C$ and $B_{\phi\rho}$. 
\begin{eqnarray}
C(z,\bar{z},\rho) &=& e^{-\rho \, L_0} \, e^{-(z-z_1) \, L_1} \, C(0) \, e^{(\bar{z}-\bar{z}_1) \, L_{-1}} \, e^{-\rho \, L_0}, \label{Cads} \\
B_{\phi\rho}(z,\bar{z},\rho) &=& e^{\rho \, L_0} \, e^{-(\bar{z}-\bar{z}_2) \, L_{-1}} \, B_{\phi\rho}(0) \, e^{(z-z_2) \, L_{1}} \, e^{\rho \, L_0}. \label{Bads}
\end{eqnarray}
Here $z_{1,2}$, $\bar{z}_{1,2}$ are complex numbers which specify the locations of  CFT operators. We checked that the components of $C \, B_{\phi\rho}$ and $B_{\phi\rho} \, C$ behave  at most as $e^{2\rho}$ for $\rho \rightarrow +\infty$. For the two operators dual to the sources $C$ and $B_{\phi\rho}$, respectively, to have same conformal weights, constant matrices $C(0)$ and  $B_{\phi\rho}(0)$ must satisfy the following pairing rule for eigenvalues of $L_0$.
\begin{eqnarray}
L_0 \, C(0)&=&-h \, C(0), \qquad C(0) \, L_0=-h' \, C(0), \nonumber \\
L_0 \, B_{\phi\rho}(0)&=&h' \, B_{\phi\rho}(0), \qquad B_{\phi\rho}(0) \, L_0 = h \, B_{\phi\rho}(0) \label{pairing}
\end{eqnarray}
In this case, $\rho$-dependence of both $C(z,\bar{z}, \rho)$ and $B_{\phi\rho}(z,\bar{z}, \rho)$ is $C(z,\bar{z},\rho) = e^{(h+h') \, \rho} \, C(e^{\rho}z,e^{\rho}\bar{z},0)$ and similar expression for $B_{\phi\rho}$. There are 9 pairs to take into account. Out of them the following three yield non-vanishing on-shell actions. 
\begin{description}
\item [(1.)] $$ C^{(1)}(0)= \left( \begin{array}{ccc} 
           1 &0&0 \\
           0 &0&0 \\
          0 &0&0 \end{array}\right),  \quad
B^{(1)}_{\phi\rho}(0)= \left( \begin{array}{ccc} 
           0 &0&0 \\
           0 &0&0 \\
          0 &0&1 \end{array}\right), \quad  (h,h')=(-1,-1) $$
\item [(2.)] $$ C^{(2)}(0)= \left( \begin{array}{ccc} 
           0 &1&0 \\
           0 &0&0 \\
          0 &0&0 \end{array}\right), \quad 
B^{(2)}_{\phi\rho}(0)= \left( \begin{array}{ccc} 
           0 &0&0 \\
           0 &0&1 \\
          0 &0&0 \end{array}\right), \quad (h,h')=(-1,0)$$
\item [(3.)] $$ C^{(3)}(0)= \left( \begin{array}{ccc} 
           0 &0&0 \\
           1 &0&0 \\
          0 &0&0 \end{array}\right), \quad 
B^{(3)}_{\phi\rho}(0)= \left( \begin{array}{ccc} 
           0 &0&0 \\
           0 &0&0 \\
          0 &1&0 \end{array}\right), \quad (h,h')=(0,-1)$$
\end{description}
The formula (\ref{Cads}) and (\ref{Bads}) give $C(z,\bar{z},\rho)$ and $B_{\phi\rho}(z,\bar{z},\rho)$, and then (\ref{renormalizedonshellaction}) gives the on-shell action. The above choices for $C(0)$ and $B(0)$ lead to the following 
on-shell action integrals. 
\begin{description}
\item [(1.)] \begin{equation}
S^{(1)}_{\text{on shell}}=\int d^2z \, \Big(-2\, \mu  \,   (z_1-z_2)(\bar{z}_1-\bar{z}_2)^2+2\bar{\mu}   \,  (z_1-z_2)^2(\bar{z}_1-\bar{z}_2)\Big) \label{S1} \end{equation}
\item [(2.)] \begin{equation} S^{(2)}_{\text{on shell}}=- \mu  \, \int d^2z \,(z_1-z_2) \label{S2} \end{equation}
\item [(3.)] \begin{equation} S^{(3)}_{\text{on shell}}=4\, \bar{\mu} \,\int d^2z \,  (\bar{z}_1-\bar{z}_2) \label{S3} \end{equation}
\end{description}
Note that the integrands do not depend on $z$ and $\bar{z}$. As mentioned at the end of sec.4, the two integrands of (\ref{renormalizedonshellaction}) do not depend on $\bar{z}$ and $z$, respectively. Furthermore, the above integrands do not depend on both coordinates, because $a_+$ and $\bar{a}_-$ are constant for AdS$_3$. 
These on-shell actions would be expected to give two-point functions. If the coefficients of both terms in $S^{(1)}_{\text{on shell}}$ were non-zero, then the two-point function does not have a suitable form, $(z_1-z_2)^{-2h} \, (\bar{z}_1-\bar{z}_2)^{-2\bar{h}}$. We must set either coefficient to zero by adjusting the parameters of the solutions; $\rho_0=2 \, \rho_1$ or $\bar{\rho}_0=2 \, \bar{\rho}_1$.  All in all, there would be four operators of conformal weights  $(h,\bar{h})=(-1,-\frac{1}{2}),  (-\frac{1}{2},-1),  (0,-\frac{1}{2})$ and $(-\frac{1}{2},0)$. 
Scalar operator with $(h,\bar{h})=(-1,-1)$, which corresponds to a Klein-Gordon scalar $\Phi=\text{tr} \, C$ \cite{HKP},  is missing. In the on-shell action $S^{(1)}_{\text{on shell}}$, the two terms have forms $\partial_{z_1} \, \big((z_1-z_2)^2 \, (\bar{z}_1-\bar{z}_2)^2 \big)$  
and $\partial_{\bar{z}_1} \, \big((z_1-z_2)^2 \, (\bar{z}_1-\bar{z}_2)^2 \big)$, respectively.   The derivative $\partial_{z_1}$ in the first term comes from an $a_{+}=L_1$ insertion in (\ref{renormalizedonshellaction}),
\begin{eqnarray}
\text{tr} \, \big( a_{+} \, b \, C \, B_{\phi\rho} \, b^{-1}   \big) &=&  \text{tr} \, \Big(L_1 \, e^{(z_1-z_2) \, L_1} \, C(0) \, e^{(\bar{z}_2-\bar{z}_1) \, L_{-1}} \, B_{\phi\rho}(0)\Big) \nonumber \\
&=&\partial_{z_1} \, \Big(\text{tr}  \, e^{(z_1-z_2) \, L_1} \, C(0) \, e^{(\bar{z}_2-\bar{z}_1) \, L_{-1}} \, B_{\phi\rho}(0)\Big).
\end{eqnarray}
This is again a result of the fact that $a_+$ is a constant matrix. 
Similarly, in the second term, $\partial_{\bar{z}_1}$ comes from $\bar{a}_{-}=-L_{-1}$ insertion. 
If there were no $a_{+}$, $\bar{a}_{-}$ insertions, then the spectrum of the conformal weights would be $(h',\bar{h}')=
(-1,-1)$, $(-1,0)$ and $ (0,-1)$. In the next section we will consider a different boundary action and obtain the scalar operator. 

Let us study the boundary conditions which these solutions satisfy. We  compute the traces of $C(z,\bar{z},\rho)$ and $B_{\phi\rho}(z,\bar{z},\rho)$ in the three cases (1.), (2.), (3.). For the first solution we have
\begin{eqnarray}
\text{tr} \, C^{(1)} &=& e^{-2\rho} \, \{1+e^{2\rho} \, (z-z_1)(\bar{z}-\bar{z}_1)\}^2, \label{C1}\\
\text{tr} \, B_{\phi\rho}^{(1)} &=& e^{-2\rho} \, \{1+e^{2\rho} \, (z-z_2)(\bar{z}-\bar{z}_2)\}^2.
\end{eqnarray}
These may be formally taken as regularization of delta function $e^{-4\rho} \, \delta^{(2)}(z-z_i)$\cite{HKP}, although the powers on the right hand sides have wrong signs to represent delta functions. 
The others are derivatives of the first one:  $\text{tr} \, C^{(2)} =-\frac{1}{2} \, \partial_{\bar{z}} \, \text{tr} \, C^{(1)}$, $\text{tr} \, C^{(3)} =-\partial_z \, \text{tr} \, C^{(1)}$. $\text{tr} \, B_{\phi\rho}^{(2)} =\frac{1}{2} \, \partial_{\bar{z}} \, \text{tr} \, B_{\phi\rho}^{(1)}$, $\text{tr} \, B_{\phi\rho}^{(3)} =\partial_z \, \text{tr} \, B_{\phi\rho}^{(1)}$. 
Because all components of $C$ are related to $\text{tr} \, C$\cite{AKP2}, all the four solutions  are connected. 
Hence it is expected that the  operators  ${\it O}^{(i)}$, $\bar{{\it O}}^{(i)}$ ($i=2,3$) are descendants of ${\it O}^{(1)}$ and $\bar{{\it O}}^{(1)}$. 
The two-point functions of these operators can be computed from $S^{(i)}_{\text{new bdry}}$ (i=2,3) separately. 

We now turn to computation of three-point correlation functions involving two operators dual to the solutions $C^{(1)}$ and $B^{(1)}$, and one higher-spin current, $<{\it O}(z_1) \bar{{\it O}}(z_2) J^{(s)}(z_3)>$. For brevity, from now on, the operator dual to $C^{(1)}$ will be denoted as ${\it O}$, and the one dual to $B^{(1)}_{\phi\rho}$ as $\bar{{\it O}}$. 

To compute the correlation functions, we need to deform the spacetime from AdS$_3$ slightly. The deformation of the metric is dual to the stress tensor, and the higher-spin gauge field the higher-spin current. Since the gauge fields $a_i$ and $\bar{a}_i$ are flat, these deformations can be achieved by gauge transformation.\footnote{The following presctiption is based on the idea in \cite{AKP2}. Here the gauge transformation (\ref{gaugeCp})-(\ref{gaugeBp}) is applied to the boundary term, not to the propagator.  } 
\begin{eqnarray}
C(z,\bar{z},\rho) & \rightarrow & b^{-1} \, U(z) \, b\, C(z,\bar{z}, \rho), \label{gaugeCp}\\
B_{\phi\rho}(z,\bar{z},\rho) & \rightarrow & B_{\phi\rho}(z,\bar{z},\rho) \, b^{-1}\, U^{-1}(z) \, b \label{gaugeBp}
\end{eqnarray}
For an infinitesimal transformation, we write $U(z)=1+\Lambda(z)$. 
However, the above transformations are slightly incorrect. Since the solutions (\ref{Csol}) and (\ref{Bsol}) are ordered exponentials, the correct transformations are,  by using (\ref{Cads}) and (\ref{Bads}), 
\begin{eqnarray}
C(z,\bar{z},\rho) & \rightarrow & e^{-\rho \, L_0} \, U(z) \, e^{-(z-z_1) \, L_1} \, U^{-1}(z_1) \, C(0) \, e^{(\bar{z}-\bar{z}_1) \, L_{-1}} \, e^{-\rho \, L_0}, \label{gaugeC} \\
B_{\phi\rho}(z,\bar{z},\rho) & \rightarrow & e^{\rho \, L_0} \, e^{-(\bar{z}-\bar{z}_2) \, L_{-1}} \, B_{\phi\rho}(0)U(z_2) \, e^{(z-z_2) \, L_{1}} \, U^{-1}(z) \, e^{\rho \, L_0}. \label{gaugeB}
\end{eqnarray}
Let us consider the first term of the on-shell action (\ref{renormalizedonshellaction}), 
\begin{equation}
S_{\text{on shell 1}} = \int d^2 z \, \text{tr} \, B_{\phi\rho} \, \partial_z \, C. 
\end{equation}
The variation of the above action to order  ${\cal O}(\Lambda^1)$ will give the three-point function.
\begin{equation}
<{\it O}(z_1) \bar{{\it O}}(z_2) J^{(s)}(z_3)> \equiv \delta \, S_{\text{on shell 1}}
\end{equation}
The variation is given by 
\begin{eqnarray}
\delta \, S_{\text{on shell 1}} &=& \int d^2z \, \text{tr} \, C(0) \, e^{(\bar{z}_2-\bar{z}_1) \, L_{-1}} B_{\phi\rho}(0)\, e^{(z-z_2) \, L_1} \, \partial_z \, \Lambda(z) \, e^{(z_1-z) \, L_1} \nonumber \\
&& + \int d^2z \, \text{tr} \, \Big\{ \big(e^{(z_1-z_2) \, L_1} \, L_1\, \Lambda(z_1)-\Lambda(z_2) \, L_1 \, e^{(z_1-z_2) \, L_1} \, \big)
\nonumber \\
&& \qquad  C(0) \, e^{(\bar{z}_2-\bar{z}_1) \, L_{-1}} \, B_{\phi\rho}(0) \Big\}. \label{deltaSonshell1}
\end{eqnarray}
The integrand of the first term depends on $z$, while that of the second term does not. 
The form of $\Lambda(z)$ is given by \cite{AKP2}\footnote{This is related to $\Lambda(\rho,z)$ in (4.16) of \cite{AKP2} by $\Lambda(z)=e^{\rho \, V^2_0} \, \Lambda(\rho,z) \, e^{-\rho \, V^2_0}$.}
\begin{equation}
\Lambda(z) = \sum_{n=1}^{2s-1} \, \frac{1}{(n-1)!} \, (-\partial_z)^{n-1} \, \Lambda^{(s)}(z) V^s_{s-n}.  \label{Lambdaz}
\end{equation}
Here $V^s_m$ is an spin-s generator of higher-spin algebra, and especially, $V^2_m=L_m$ and $V^3_m=W_m$ \footnote{For $W_m$, see appendix A.}  in the case of $sl(3,R)$. 
In CFT the three-point function $<{\it O}(z_1) \bar{{\it O}}(z_2) J^{(s)}(z_3)>$ of primary operators ${\it O}$ and $\bar{{\it O}}$ must have a form 
\begin{equation}
<{\it O}(z_1) \bar{{\it O}}(z_2) J^{(s)}(z_3)> \propto \Big(\frac{z_1-z_2}{(z_1-z_3)(z_2-z_3)}\Big)^s \, <{\it O}(z_1) \tilde{{\it O}}(z_2) >.  \label{OOJ}
\end{equation}
We consider the case of spin 2 (s=2) transformation, and set $\Lambda^{(2)}(z)=\frac{1}{2\pi(z-z_3)}$. 
When $C^{(1)}$ and $B^{(1)}_{\phi\rho}$ for the solution (1.) is substituted into the second term of (\ref{deltaSonshell1}), we have 
\begin{eqnarray}
\int d^2z \, \frac{(z_1-z_2)(\bar{z}_1-\bar{z}_2)^2}{\pi (z_1-z_3)^2(z_2-z_3)^2} \, \{ (z_1-z_3)^2+(z_2-z_3)^2-(z_1-z_3)(z_2-z_3) \}.
\end{eqnarray} 
Since  $<{\it O}(z_1) \bar{{\it O}}(z_2) >= (z_1-z_2)(\bar{z}_1-\bar{z}_2)^2$, this does not agree with (\ref{OOJ}) with $s=2$. There also exists the extra first term in (\ref{deltaSonshell1}). Hence the correlation functions in (\ref{S1}) are not those of primary operators.  Similar analysis to the cases (2.)-(3.) shows that the correlation functions (\ref{S2})-(\ref{S3}) are not, either.

\section{New Local Boundary Term}
\hspace{5mm}
With the observation in the previous section  we are forced to set the on-shell action (\ref{renormalizedonshellaction}) to zero by adjusting the parameters of the local counterterms in such a way that $\mu=\bar{\mu}=0$.  Then the residual gauge symmetry (\ref{residualsym}) is recovered.  In order to obtain the generating functional of the correlation functions of CFT operators, we need to introduce new appropriate boundary terms. In this context the example\cite{Hen}\cite{LR} of a free spinor is helpful. In AdS/CFT correspondence for a free spinor $\psi$, a boundary term $\int_{\partial {\cal M}} d^2x \, \sqrt{\gamma} \, \bar{\psi}\psi$ is added to the bulk action,\footnote{$\gamma_{ij}$ is an induced metric on $\partial {\cal M} $.} because the bulk action vanishes when a solution to the equation of motion is substituted. This fermion boundary term keeps all the symmetry required. 

A simplest prescription would be to adopt the surface term, which appears in the action of a free scalar field in AdS background, after substitution of the solution to the equation of motion and partial integration. 
Since $\text{tr} \, C$ obeys Klein-Gordon equation with mass $m^2=\lambda^2-1$ with $\lambda=3$\cite{AKP2},   a term like  
$\int_{\partial {\cal M}} d^2x \, \sqrt{\gamma} \ \text{tr} B_{\phi\rho} \, \text{tr} \, C$, 
where $\gamma_{\mu\nu}=e^a_{\mu} \, e_{a\nu}|_{\partial M}$ is the induced metric on the boundary, is expected to work. Here, a derivative on $\text{tr} \, C$ with respect to $\rho$ is not introduced.  Such a derivative will simply modify the generating functional by a multiplicative constant.  It can be shown that this surface term give an appropriate generating functional for two-point functions. In the case of HS[$\frac{1}{2}$] $\times$ HS[$\frac{1}{2}$]
higher-spin gravity which will be treated in sec.7, however, this prescription for calculating two-point functions would give those for operators of scaling dimension $\Delta_+=\frac{3}{2}$ or $\Delta_-=\frac{1}{2}$ only. Moreover, this boundary term breaks the residual gauge symmetry (\ref{residualsym}). Hence, in the remainder of this paper, we will not exploit this boundary term.  

We propose the following new local boundary term.
\begin{equation}
S_{\text{new bdry}} = \lim_{\rho \rightarrow \infty} \int d^2z \, \frac{g}{{\cal A}} \, \text{tr} \, \big(B_{\phi\rho} (z,\bar{z},\rho) \, C(z,\bar{z},\rho)\big) \label{newS}
\end{equation}
Here $g$ is a constant different from zero. ${\cal A}$ is the area of the boundary.  This boundary term will be added to the action integral from the beginning. Addition of a new boundary term modifies the theory. The relative coefficient between the term (\ref{newS}) and the other part of the action (\ref{SCS})+(\ref{SBC})+ (\ref{surface}) is not determined in the present context, and we will simply set the coefficient of (\ref{newS}) to a non-vanishing constant $g$. The above term (\ref{newS})  is invariant under the residual gauge transformation (\ref{residualsym}).  When the solution to the equations of motion is substituted, this trace is independent of $\rho$.\footnote{In the case of a Klein-Gordon scalar field $\phi(x)$ in AdS background, the integrand of the boundary action which appears after partial integration, $\phi(x) \, \partial_{\rho} \, \phi(x)$, depends on $\rho$.}  Hence actually, the limit $\rho \rightarrow \infty$ is not necessary. The reason for this peculiar phenomenon is that the equation of motion for $C$ makes the solution a parallel transport of its value at a single point, and that (\ref{newS}) is gauge invariant. The data of the fields are transfered to the internal space. 
Hence the  term (\ref{newS}) captures the property of the matter fields more properly. 
It will be shown that the above boundary term (\ref{newS}) works as a correct generating functional for two-point function and three-point functions. The integrand is also independent of  $z$ and $\bar{z}$, so that the area ${\cal A}$ of the boundary will be cancelled out. So (\ref{newS}) can be replaced by  $S_{\text{new bdry}} = g \, \text{tr} \, \big(B_{\phi\rho} (z,\bar{z},\rho) \, C(z,\bar{z},\rho)\big)$. 

When the solution (1.) in sec.5 is substituted into (\ref{newS}), we obtain the following on-shell action. 
\begin{eqnarray}
S^{(1)}_{\text{new bdry}} &=& g \, (z_1-z_2)^2(\bar{z}_1-\bar{z}_2)^2
\end{eqnarray}
The two-point function of the operators ${\it O}$ and $\bar{{\it O}}$ dual to $C^{(1)}$ and $B^{(1)}$ are given by $<{\it O}(z_1) \, \bar{{\it O}}(z_2)>=(z_1-z_2)^2(\bar{z}_1-\bar{z}_2)^2$ and the conformal weights $(h,\bar{h})$ are $(-1,-1)$. 

Due to the transformations (\ref{gaugeC}), (\ref{gaugeB}) of $C$ and $B_{\phi\rho}$ with extra variations at $z_1$ and $z_2$, the `invariant' boundary action (\ref{newS}) transforms as 
\begin{equation}
\delta \, S_{\text{new bdry}} = g \, \text{tr} \,  \Big\{ e^{(\bar{z}_2-\bar{z}_1) \, L_{-1}} \, B_{\phi\rho}(0) \, \bigg( \Lambda(z_2) \, e^{(z_1-z_2) \, L_1}-e^{(z_1-z_2) \, L_1} \, \Lambda(z_1) \bigg)\,C(0) \Big\}.
\end{equation}
For spin-2 transformation with $\Lambda(z)=\frac{1}{2\pi}(\frac{1}{z-z_3} \, L_1+\frac{1}{(z-z_3)^2} \, L_0+\frac{1}{(z-z_3)^3} \, L_{-1})$, the variation of the action with the solution (1.) substituted, is given by 
\begin{eqnarray}
\delta \, S^{(1)}_{\text{new bdry}} &=& -\frac{g}{2\pi} \, \Big(\frac{z_1-z_2}{(z_1-z_3)(z_2-z_3)}\Big)^2 \, (z_1-z_2)^2(\bar{z}_1-\bar{z}_2)^2.
\end{eqnarray}
Hence the solution  satisfies the relation (\ref{OOJ}). 
For spin-3 transformation with $\Lambda(z)=\frac{1}{2\pi}(\frac{1}{z-z_3} \, W_2+\frac{1}{(z-z_3)^2} \, W_1+\frac{1}{(z-z_3)^3} \, W_0+\frac{1}{(z-z_3)^4} \, W_{-1}+\frac{1}{(z-z_3)^5} \, W_{-2})$, we obtain
\begin{eqnarray}
\delta \, S^{(1)}_{\text{new bdry}} &=& \frac{1}{3\pi} \, \Big(\frac{z_1-z_2}{(z_1-z_3)(z_2-z_3)}\Big)^3 \, S^{(1)}_{\text{new bdry}}.
\end{eqnarray}
Hence the operators ${\it O}$ and $\bar{{\it O}}$ dual to $C^{(1)}$ and $B^{(1)}_{\phi\rho}$ are primary ones.

\section{Matter Fields Coupled to $HS[\frac{1}{2}] \times HS[\frac{1}{2}]$ CS theory}
\hspace{5mm}
The above analysis can be extended to the 3d higher-spin gravity based on $HS[\lambda] \times HS[\lambda]$ gauge symmetry.\cite{KP2} In the action integral $S_{CS-BC}$, product of matrices $A_{\mu}$, $\bar{A}_{\mu}$, $C$ and $B_{\mu\nu}$ must be simply replaced by their lone-star product $\star$\cite{lonestar}\cite{AKP2}. When the parameter $\lambda$ is equal to $\frac{1}{2}$, this product reduces to a Moyal product $*$, and the calculation simplifies.\cite{AKP2}  In what follows, we will restrict discussion only to this case. The Moyal product of two functions  is defined by 
\begin{equation}
(f*g)(y) = \frac{1}{4\pi^2} \, \int d^2u \, d^2v \, f(y+u) \, g(y+v) \, e^{iuv}.
\end{equation}
Here $y_{\alpha}$, $u_{\alpha}$ and $v_{\alpha}$ $(\alpha=1,2)$ are twister variables, with $uv=\epsilon^{\alpha\beta} \, u_{\alpha} \, v_{\beta}$, and $\epsilon^{12}=1$. The $*$-commutator of $y_{\alpha}$ is given by $[y_1,y_2]_*=2i$. The generators of $hs[\frac{1}{2}]$ are defined by even polynomials of $y_{\alpha}$. 
\begin{equation}
V^s_m= \left(\frac{-i}{4}\right)^{s-1} \, y_1^{s+m-1} \, y_2^{s-m-1}. \qquad (s=2,3,\dots; 1-s \leq m \leq s-1) \label{Vsm}
\end{equation}
Especially, s=2 generators satisfy the algebra of $L_i$'s: $[V^2_1, V^2_0]_*=V^2_1$, $[V^2_1,V^2_{-1}]_*=2V^2_0$ and $[V^2_0,V^2_{-1}]_*=V^2_{-1}$. The trace operation is replaced by 
\begin{equation}
\text{tr}_y \, f(y_1,y_2)= f(0,0).
\end{equation}

In this section the matter fields are extended to include odd polynomials of $y_{\alpha}$. They are split as
\begin{equation}
C=C^{\text{e}}+C^{\text{o}}, \qquad B_{\phi\rho}=B_{\phi\rho}^{\text{e}} + B_{\phi\rho}^{\text{o}},\label{eo}
\end{equation}
where the fields with a superscript e or o are made of even and odd polynomials, respectively.  $C^{\text{e}}$ and $B_{\phi\rho}^{\text{e}}$ are bosons. $C^{\text{o}}$ and $B_{\phi\rho}^{\text{o}}$ represent fermionic fields, since $\text{tr}_y \, B_{\phi\rho}^{\text{o}} \, C^{\text{o}}=- \text{tr}_y \,C^{\text{o}} \, B_{\phi\rho}^{\text{o}}$, due to a formula
\begin{equation}
\text{tr}_y \, f(y) * g(y)=\text{tr}_y \, g(-y)*f(y)=\text{tr}_y \, g(y) * f(-y). \label{tracerule}
\end{equation}
When (\ref{eo}) is substituted into the action (\ref{SBC}), the action for the bosonic fields and the fermionic ones decouple, because Moyal product keeps parity of polynomials and the trace of odd polynomials vanish.

\subsection{New Eigenfunctions of $V^2_0$}
\hspace{5mm}
The eigenfunctions of $V^2_0$ are given by\cite{KP2} 
\begin{equation}
f_{mn}(y) \equiv y_1^m * e^{-iy_1y_2} * y_2^n. \qquad (m,n=0,1, \dots) \label{oldeigen}
\end{equation}
These functions satisfy
\begin{eqnarray}
V^2_0 * f_{mn} (y) &=&-\frac{2m+1}{4} \, f_{mn}(y), \nonumber \\
f_{mn}(y)  * V^2_0 &=& -\frac{2n+1}{4} \, f_{mn}(y). \label{V20eigen}
\end{eqnarray}
This fact (for $m=n=0,1$) was used in \cite{KP2} to construct two scalar functions $\text{Tr} \, C_{\pm}$.  
However, in this case the eigenvalues of (\ref{V20eigen}) are all negative. This does not fit to our purpose. 

Let us define another set of  eigenfunctions $\tilde{f}_{mn}(y)$ by
\begin{equation}
\tilde{f}_{mn}(y) \equiv y_2^m * e^{iy_1y_2} * y_1^n. \qquad (m,n=0,1, \dots) \label{neweigen}
\end{equation}
It can be shown that this satisfies the equations.  
\begin{eqnarray}
V^2_0 * \tilde{f}_{mn} (y) &=&\frac{2m+1}{4} \, \tilde{f}_{mn}(y), \nonumber \\
\tilde{f}_{mn}(y)  * V^2_0 &=& \frac{2n+1}{4} \, \tilde{f}_{mn}(y). \label{V20eigen2}
\end{eqnarray}
(\ref{oldeigen}) and (\ref{neweigen}) allow us to define pairs of $C(0)$ and $B_{\phi\rho}(0)$ that obey (\ref{pairing}). There are two types 
of assignments of eigenfunctions $f_{mn}$ and $\tilde{f}_{mn}$. 
\begin{description}
\item [A.]
\begin{equation}
C^{(m,n)}(0) = f_{mn}(y), \qquad B^{(n,m)}_{\phi\rho}(0)= \tilde{f}_{nm}(y). \qquad (m,n =0,1, \dots)
\end{equation}
\item  [B.]
\begin{equation}
C^{(m,n)}(0) = \tilde{f}_{mn}(y), \qquad B^{(n,m)}_{\phi\rho}(0)= f_{nm}(y). \qquad (m,n =0,1, \dots)
\end{equation}
\end{description}

\subsection{Two-Point Functions}
\hspace{5mm}
The new boundary action (\ref{newS}) for this model is given by the following two traces. 
\begin{eqnarray}
S^{(A,mn)}_{\text{new bdry}} &=&  g \, \text{tr}_y \, \Big( y_2^{n}*e^{iy_1y_2}*y_1^{m} *  e^{-\frac{i}{4}(z_1-z_2)y_1^2}   \nonumber \\
&&
* y_1^m * e^{-iy_1y_2}* y_2^n* e^{\frac{i}{4} (\bar{z}_1-\bar{z}_2) y_2^2} \Big), \label{SnbA} \\
S^{(B,mn)}_{\text{new bdry}} &=&  g \, \text{tr}_y \, \Big(y_1^{n}*e^{-iy_1y_2}*y_2^{m} * e^{-\frac{i}{4}(z_1-z_2)y_1^2}  \nonumber \\
&& * y_2^{m} * e^{iy_1y_2} * y_1^{n} * e^{\frac{i}{4} (\bar{z}_1-\bar{z}_2) y_2^2}
 \Big)
\end{eqnarray}
These on-shell actions can be computed by using the method in \cite{KP2}.  Some new formulae are 
presented in appendix B. By differentiating (\ref{prodmp}) with respect to $z_1$ and $\bar{z}_1$ necessary 
times, we obtain $S^{(A,m,n)}_{\text{new bdry}} $. 
\begin{eqnarray}
 \nonumber \\
 S^{(A,mn)}_{\text{new bdry}}&=& \frac{1}{2} \, g \, (-i)^{n+m} \, \frac{(2m)! \, (2n)!}{m! \, n!} \,   \, (z_1-z_2)^{-\frac{1}{2}-m} \, (\bar{z}_1-\bar{z}_2)^{-\frac{1}{2}-n} \label{hs2pt}
\end{eqnarray}
Here  formula (\ref{tracerule}) is used to move $y_2^n$ in the middle of the trace to the leftmost. Note that the limit $\rho \rightarrow \infty$ is not taken.

The second surface on-shell action $ S^{(B,mn)}_{\text{new bdry}}  $ is ill-defined: by using (\ref{B2}) one obtains
\begin{multline}
\text{tr}_y \, e^{-\frac{i}{4} \, z y_1^2} * e^{\xi_2y_2}* e^{iy_1y_2}*e^{\eta_1y_1}*e^{\frac{i}{4} \, \bar{z}y_2^2}*e^{\xi_1y_1}
*e^{-iy_1y_2}*e^{\eta_2y_2}  \\
=e^{2i(\xi_1\eta_2-\xi_2\eta_1)+iz(\xi_2)^2-i\bar{z}(\xi_1)^2} \, \text{tr}_y \, e^{iy_1y_2+2(\eta_1y_1+\xi_2y_2)} * e^{-iy_1y_2+2(\xi_1y_1+\eta_2y_2)}
\end{multline}
Here $\eta_i$ and $\xi_i$ are sources for $y_i$, and $z=z_1-z_2$, $\bar{z}=\bar{z}_1-\bar{z}_2$. When (\ref{B2}) is again applied to the Moyal product in the last line, one obtains a divergent result. 
This divergence cannot be removed by an overall renormalization of the solutions. Even if this divergence is regularized temporally, it does not lead to conformally covariant two-point functions. 
This asymmetry in $C$ and $B$ occurs, because in (\ref{asmAdS3}),  $a_+$ is associated with $L_1$ and $\bar{a}_-$ with $L_{-1}$. 
Similarly, in the spin-3 case in secs.4 and 5, the on-shell action completely vanishes for $C$ being a right-lower triangular matrix and $B$ a left-upper triangular one. We will not consider this type of solutions. 

\subsection{Primary Operators}
\hspace{5mm}
Let us now identify primary operators. The A-type solutions $C^{(m,n)}(z,\bar{z},\rho) =e_*^{-\rho \, V^2_0}*e^{-(z-z_1) \, V^2_1} *f_{mn}(y)*e^{(\bar{z}-\bar{z}_1) \, V^2_{-1}} *e_*^{-\rho \, V^2_0}$ can be classified into four sets according to the values of $m$ and $n$ (mod  $2\mathbb{Z})$. 
\begin{multline}
 C^{(\nu_1+2a,\nu_2+2b)}(z,\bar{z},\rho)  \\
=   e_*^{-\rho \, V^2_0}*e^{\frac{i}{4}(z-z_1) \, y^2_1} *y_1^{\nu_1+2a}* e^{-iy_1y_2}* y_2^{\nu_2+2b}*e^{-\frac{i}{4}(\bar{z}-\bar{z}_1) \, y^2_{2}} *e_*^{-\rho \, V^2_0}  \\
= e^{(h+\bar{h})  \rho}  \, \Big( e^{\frac{i}{4} \, (z-z_1)e^{\rho}y_1^2} *y_1^{\nu_1+2a}* e^{-iy_1y_2}* y_2^{\nu_2+2b}*
e^{-\frac{i}{4}e^{\rho}(\bar{z}-\bar{z}_1) \, y^2_{2}}\Big). \label{Czzr}
\end{multline}
Here $\nu_1, \nu_2=0,1$ and $a,b=0,1,2,\cdots$ and 
\begin{equation}
(h,\bar{h}) =\Big(\frac{2\nu_1+1}{4}+a, \frac{2\nu_2+1}{4}+b \Big)
\end{equation}
is the conformal weight. Since $y_1^{2a}$ and $y_2^{2b}$ can be replaced by $(\partial_z)^a$ and $(\partial_{\bar{z}})^b$, respectively, up to multiplying constants, 
the solutions with $a\geq 1$ or $b \geq 1$ correspond to descendants.  There are four solutions $C^{(\nu_1,\nu_2)}$ which are dual to the primary operators. 

Solutions $C^{(0,0)}$ and $C^{(1,1)}$ are already obtained in \cite{KP2} and denoted as $C_-$ and $2iC_+$. 
\begin{eqnarray}
C^{(0,0)}(z,\bar{z},\rho;z_1,\bar{z}_1) &=& e^{\frac{1}{2} \, \rho} \, \frac{1}{\sqrt{|L|}} \, e^{\frac{1}{2} \, y^T \, S \, y},  \label{C00sol}\\
C^{(1,1)}(z,\bar{z},\rho;z_1,\bar{z}_1) &=& e^{\frac{3}{2} \, \rho} \, \frac{1}{\sqrt{|L|}} \, e^{\frac{1}{2} \, y^T \, S \, y}  \Big\{ \frac{2i}{|L|} \nonumber \\
&&+\frac{4}{|L|^2} \, (y_1+(\bar{z}-\bar{z}_1)e^{\rho}y_2) \, (y_2-(z-z_1)e^{\rho}y_1)\Big\}
\end{eqnarray}
Here $|L|$ and $S$ have the following expressions. 
\begin{eqnarray}
|L| &=& e^{2\rho} \, |z-z_1|^2+1,  \\
S  &=&\frac{i}{|L|} \, \begin{pmatrix} 2e^{\rho} \,  (z -z_1)  & |L|-2 \\  |L|-2 & -2 e^{\rho} \, (\bar{z}-\bar{z}_1) \end{pmatrix}.      
\end{eqnarray}
The traces of these fields formally behave for $\rho \rightarrow \infty$ as 
\begin{eqnarray}
\text{tr}_y \, C^{(0,0)} & \sim & -2\pi e^{-\frac{3}{2} \, \rho} \, \delta^{(2)}(z-z_1)+e^{-\frac{1}{2} \, \rho} \, |z-z_1|^{-1}, \label{C00asym}\\
\frac{1}{2i} \, \text{tr}_y \, C^{(1,1)} & \sim & 2\pi \, e^{-\frac{1}{2} \, \rho} \, \delta^{(2)}(z-z_1)+ e^{-\frac{3}{2} \, \rho} \, |z-z_1|^{-3}. \label{C11asym}
\end{eqnarray}
In analogy with the usual AdS/CFT correspondence for a scalar field $\phi(x)$ \cite{ads3}\cite{ads4}, these traces are expected to be sources of scalar operators in CFT on the boundary. There is, however, a difference between the BC model coupled to higher spin gravity and the ordinary scalar field theory. It is known that there are two ways to quantize a scalar field in AdS \cite{BF}, and there are corresponding operators ${\cal O}_+$ and ${\it O}_-$.  In the usual AdS/CFT correspondence, the scalar field $\phi$ works as a source for ${\it O}_+$, and the generating functional of correlation functions of ${\it O}_-$ is obtained by Legendre transformation of that of ${\it O}_+$. 
In the present case, operators dual to $\text{tr} \, C^{(0,0)}$ and $\frac{1}{2i} \, \text{tr} \, C^{(1,1)}$ are denoted as
${\it O}^{(0,0)}$ and ${\it O}^{(1,1)}$, respectively. Similarly, operators  dual to $\text{tr} \, B_{\phi\rho}^{(0,0)}$ and  $\frac{1}{2i} \, \text{tr} \, B_{\phi\rho}^{(1,1)}$ are denoted as $\bar{{\it O}}^{(0,0)}$ and $\bar{{\it O}}^{(1,1)}$. 
Then the two-point correlation functions of these operators are obtained from (\ref{hs2pt}) straightforwardly (without $\rho \rightarrow \infty$ limit and Legendre transformation)  as 
\begin{eqnarray}
<{\it O}^{(0,0)}(z_1) \, \bar{{\it O}}^{(0,0)}(z_2)> &=& \frac{1}{2} \, g \, (z_1-z_2)^{-1/2}(\bar{z}_1-\bar{z}_2)^{-1/2}, \label{O00O00}\\
<{\it O}^{(1,1)}(z_1) \, \bar{{\it O}}^{(1,1)}(z_2)> &=& \frac{1}{2} \, g \, (z_1-z_2)^{-3/2}(\bar{z}_1-\bar{z}_2)^{-3/2}.
\end{eqnarray}
This point is in sharp contrast to the usual AdS/CFT correspondence in a free scalar theory\cite{ads4}.  All primary operators can be quantized in a single quantization.\footnote{For off-diagonal two-point functions, see the end of this subsection.}

Solutions  $C^{(0,1)}$ and $C^{(1,0)}$  do not belong to  $hs[\frac{1}{2}]$, because  
\begin{eqnarray}
C^{(0,1)}(z,\bar{z},\rho;z_1,\bar{z}_1) &=& e^{ \rho} \, \frac{2}{|L|^{\frac{3}{2}}} \, e^{\frac{1}{2} \, y^T \, S \, y}  \, \big ( y_2- e^{\rho} \, (z-z_1) \, y_1\big),  \\
C^{(1,0)}(z,\bar{z},\rho;z_1,\bar{z}_1) &=& e^{ \rho} \,  \frac{2}{|L|^{\frac{3}{2}}} \, e^{\frac{1}{2} \, y^T \, S \, y}   \, \big (y_1+ e^{\rho} \, (\bar{z}-\bar{z}_1) \, y_2\big). 
\end{eqnarray}
They are made up of terms with odd number of $y$'s, and are fermions. These have spinor components. 
\begin{gather}
\text{tr}_y \, C^{(0,1)}*y_2 = \frac{-2i}{|L|^{\frac{3}{2}}} \, e^{2\rho} \, (z-z_1), \quad
\text{tr}_y \, C^{(0,1)}*y_1 = \frac{-2i}{|L|^{\frac{3}{2}}} \, e^{\rho}, \label{spinor} \\
\text{tr}_y \, C^{(1,0)}*y_1 = \frac{-2i}{|L|^{\frac{3}{2}}} \, e^{2\rho} \, (\bar{z}-\bar{z}_1), \quad
\text{tr}_y \, C^{(1,0)}*y_2 = \frac{2i}{|L|^{\frac{3}{2}}} \, e^{\rho}. \label{spinor2}
\end{gather}
The components of  $C^{(01)}$ behave for $\rho \rightarrow \infty$ as
\begin{eqnarray}
\text{tr}_y \,  C^{(01)} * y_2 & \sim & -8\pi i \, e^{-2\rho} \, \partial_{\bar{z}} \, \delta^{(2)}(z-z_1)-2i \, e^{-\rho} \, (z-z_1)^{-\frac{1}{2}} \, (\bar{z}-\bar{z}_1)^{-\frac{3}{2}}, \nonumber  \\
\text{tr}_y
 \, C^{(01)} * y_1 & \sim & -4\pi i \,  e^{-\rho} \, \delta^{(2)}(z-z_1)-2i \, e^{-2\rho} \, (z-z_1)^{-\frac{3}{2}} \, (\bar{z}-\bar{z}_1)^{-\frac{3}{2}}. \label{C01two}
\end{eqnarray}
Solution to the equation of motion for a free massive Dirac fermion in AdS$_3$ with a chiral boundary condition is given by\cite{Hen}\cite{LR}
\begin{equation}
\psi(\bm{x},\rho)  = \int d^2\bm{x}_1 \, \Big(e^{-\rho}\,\Gamma^0   +(\bm{x}-\bm{x}_1) \cdot \bm{\Gamma}\Big) \, \Big(e^{-2\rho}+|\bm{x}-\bm{x}_1|^2\Big)^{-\frac{3}{2}+m\Gamma^0} \, e^{\rho(-1+m\Gamma^0)} \, \psi_0(\bm{x}_1). \label{solHen}
\end{equation}
Here $\Gamma^0=-i\Gamma^1\Gamma^2$ is a gamma matrix in the radial direction, and $\bm{\Gamma}=(\Gamma^1,\Gamma^2)$, those in the $\phi, t$ directions. For a chiral boundary condition which imposes $\Gamma^0 \, \psi_0=+\psi_0$,  one has $(\bm{x}-\bm{x}_1) \cdot \bm{\Gamma} \, \psi_0=(z-z_1) \, \Gamma^1 \, \psi_0$. 
By comparing this expression (\ref{solHen}) with the two components (\ref{spinor}), one can identify, if one sets $m=0$,
\begin{equation}
\psi(z,\bar{z},\rho)=\frac{i}{2} \, \int d^2z_1 \,  \text{tr}_y \, (C^{(01)}(z,\bar{z},\rho) * (y_1-y_2 \Gamma^1) ) \, \psi_0(z_1,\bar{z}_1).  
\end{equation}
The components of $C^{(10)} $ (\ref{spinor2}) can also be related to anti-chiral spinor $\psi_0$ with $\Gamma^0 \, \psi_0=-\psi_0$. 
The operators dual to $\text{tr} \, (C^{(01)} * (y_1-y_2 \, \Gamma^1)) $  and  $\text{tr} \, (C^{(10)} * (y_2+ y_1 \, \Gamma^1)) $ will be denoted as ${\it O}^{(0,1)}$ and  ${\it O}^{(1,0)}$, respectively.  The solutions $B_{\phi\rho}^{(\nu_1,\nu_2)}$ can also be computed, 
 and $\bar{{\it O}}^{(\nu_1,\nu_2)}$ are defined similarly.  
Two-point correlation functions of these operators are obtained from (\ref{hs2pt}) as 
\begin{eqnarray}
<{\it O}^{(0,1)}(z_1) \, \bar{{\it O}}^{(1,0)}(z_2)> &=& -i \, g \, (z_1-z_2)^{-1/2}(\bar{z}_1-\bar{z}_2)^{-3/2}, \nonumber \\
<{\it O}^{(1,0)}(z_1) \, \bar{{\it O}}^{(0,1)}(z_2)> &=&- i \, g \, (z_1-z_2)^{-3/2}(\bar{z}_1-\bar{z}_2)^{-1/2}. \label{fermioncorrel}
\end{eqnarray}
They  have conformal weights, $(h,\bar{h})=(\frac{1}{4}, \frac{3}{4})$ and $(\frac{3}{4},\frac{1}{4})$, respectively. These have spin 
$h-\bar{h}=\pm \frac{1}{2}$. The scaling dimension is given by $\Delta \equiv h+\bar{h}=1$. Comparing this with the result of \cite{Hen},  $\Delta=\frac{d}{2}+m$ with $d=2$, we again find  that the fermion is massless, $m=0$. 

One can show that off-diagonal two-point functions vanish: let us take linear combinations of the independent solutions $C^{(0,0)}$ and $C^{(1,1)}$ in (\ref{C00asym}) and (\ref{C11asym}).
\begin{multline}
C(z,\bar{z},\rho) = -\frac{1}{2\pi} \, \int d^2z_1 \, C^{(0,0)}(z,\bar{z},\rho;z_1,\bar{z}_1) \, \phi_0(z_1,\bar{z}_1) \\-\frac{1}{4\pi i} \, 
\int d^2z_1 \, C^{(1,1)}(z,\bar{z},\rho;z_1,\bar{z}_1) \, \phi_1(z_1,\bar{z}_1)
\end{multline}
Here $\phi_0(z_1,\bar{z}_1)$ and $\phi_1(z_1,\bar{z}_1)$ are boundary conditions for the two modes. Similarly for $B_{\phi\rho}$\footnote{$B_{\phi\rho}^{(0,0)}(z,\bar{z},\rho;z_2,\bar{z}_2) $ and $B_{\phi\rho}^{(0,0)}(z,\bar{z},\rho;z_2,\bar{z}_2) $ are 
the counterparts of (\ref{C00asym}) and (\ref{C11asym}). }:
\begin{multline}
B_{\phi\rho}(z,\bar{z},\rho) = -\frac{1}{2\pi} \, \int d^2z_2 \, B_{\phi\rho}^{(0,0)}(z,\bar{z},\rho;z_2,\bar{z}_2) \, \bar{\phi}_0(z_2,\bar{z}_2) \\-\frac{1}{4\pi i} \, 
\int d^2z_2 \, B_{\phi\rho}^{(1,1)}(z,\bar{z},\rho;z_2,\bar{z}_2) \, \bar{\phi}_1(z_2,\bar{z}_2)
\end{multline}
Then the boundary term $\text{tr} \, B_{\phi\rho} \, C$ is given by
\begin{multline}
\frac{1}{8\pi^2} \, \int d^2z_1 \, \int d^2z_2 \, \phi_0(z_1,\bar{z}_1) \, \bar{\phi}_0(z_2,\bar{z}_2) \, \frac{1}{|z_1-z_2|} \\
+\frac{1}{8\pi^2} \, \int d^2z_1 \, \int d^2z_2 \, \phi_1(z_1,\bar{z}_1) \, \bar{\phi}_1(z_2,\bar{z}_2) \, \frac{1}{|z_1-z_2|^3}
\end{multline}
We also obtain a similar result for fermionic solutions, and we conclude that
\begin{equation}
<{\it O}^{(\nu_1,\nu_2)}(z_1,\bar{z}_1) \, \bar{{\it O}}^{(\nu_2',\nu_1')}(z_2,\bar{z}_2)>=0, \qquad \text{if   } (\nu_1,\nu_2) \neq (\nu_1',\nu_2').
\end{equation}
Hence, in our prescription we can consider all the primary operators, ${\it O}^{(\nu_1,\nu_2)}$ and $\bar{{\it O}}^{(\nu_2,\nu_1)}$, in a single quantization. In the usual AdS/CFT correspondence of the scalar field\cite{ads3}\cite{ads4}, only operator ${\it O}^{(1,1)}$ can be considered in the standard quantization, and only ${\it O}^{(0,0)}$ in the alternate quantization. In the minimal model holography\cite{GG},  two operators ${\it O}^{(0,0)}$ and ${\it O}^{(1,1)}$ are expected to work as ladder operators to generate other primary operators by fusion, and two complex scalar fields are introduced in \cite{GG}.  In our formalism it is possible to introduce both primary operators by means of  a single field $C$.\footnote{If two sets of fields $C$ and $B$ were introduced, then the symmetry transformation (\ref{extralocalsymmetry}) could not be extended to the two sets.} It would be interesting if there were a formalism, where in a theory of a single real bulk scalar field $\phi$ of mass $-\frac{1}{4} d^2 < m^2 < 1-\frac{1}{4} \, d^2$ in AdS$_{d+1}$, both operators of scaling dimensions $\Delta_{\pm}= \frac{1}{2} \, d \pm \sqrt{\frac{1}{4} \, d^2+m^2}$ appear in a boundary CFT. Actually, there are several quantizations for scalar fields in AdS.\cite{Avis} Difficulty, however, lies in the difference of the relative signs of the two terms in (\ref{C00asym}) and (\ref{C11asym}). This might lead to non-unitarity of CFT. In the calculation of the two-point functions presented in this paper this problem does not occur, because the asymptotic behaviors (\ref{C00asym}) and (\ref{C11asym}) are not used.

In appendix C, two scalar propagators on BTZ black hole\cite{BTZ} at $\lambda=\frac{1}{2}$ are studied and it is shown that the results in the literature\cite{KP2} are obtained by means of our boundary action (\ref{newS}). Furthermore, two-point functions of operators with different scaling dimensions vanish. Hence our prescription also works in backgrounds more complicated than AdS vacuum. 

\subsection{Three-Point Correlation Functions}
\hspace{5mm}
We will compute the three-point functions in order to check the operators discussed in the previous 
subsection are not just quasi-primary, but really primary ones. 
These correlation functions are obtained by gauge transformation of the boundary action 
$S^{A, \nu_1, \nu_2}_{\text{new bdry}}$. 
\begin{eqnarray}
&& <{\it O}^{(\nu_1,\nu_2)}(z_1) \, \bar{{\it O}}^{(\nu_2,\nu_1)}(z_2) \, J^{(s)}(z_3)> \label{3point} \\
&=& g \, \text{tr}_y \, \Big\{  e^{(\bar{z}_2-\bar{z}_1) \, V^2_{-1}} \, * \tilde{f}_{\nu_2,\nu_1} * 
(\Lambda(z_2) * e^{(z_1-z_2) V^2_1}-e^{(z_1-z_2) V^2_1}* \Lambda(z_1))*  f_{\nu_1,\nu_2}\Big\} \nonumber 
\end{eqnarray}
The gauge parameter $\Lambda(z)$ is given by (\ref{Lambdaz}) with $\Lambda^{(s)}=\frac{1}{2\pi(z-z_3)}$. 

We will explain the procedure by the case of $(\nu_1,\nu_2)=(0, 0)$. By expressing $V^s_m$ in $\Lambda(z)$ in terms of 
$y_1^a \, y_2^b$ by using (\ref{Vsm}), and evaluating the trace in (\ref{3point}) for spin-s transformation, we have
\begin{eqnarray}
\Big(\frac{-i}{4}\Big)^{s-1} \, \frac{-1}{4\pi\sqrt{z_{12}\bar{z}_{12}}} \, \sum_{n=1}^{2s-1} \, \frac{(2s-n-1)!}{(s-n)!} \, \Big ( (-i)^{n-1}\frac{1}{z^n_{13}} \, \Big(\frac{-i}{z_{12}}\Big)^{s-n} - (i)^{n-1}\frac{1}{z^n_{23}} \, \Big(\frac{-i}{z_{12}}\Big)^{s-n} \Big) \label{threep00}
\end{eqnarray}
Here $z_{ij}=z_i-z_j$, {\em etc}. This summation can be performed by using the following identity.
\begin{eqnarray}
\sum_{n=1}^s \frac{(2s-n-1)!}{(s-n)!} \, x^n &=&  \frac{(2s-2)!}{(s-1)!} \, \frac{2x}{2-x} \, _2F_1 \left( 
\frac{1}{2}, 1,\frac{3}{2}-s;\Big( \frac{x}{2-x}\Big)^2 \right) \nonumber \\
&& + \frac{1}{2} \, (-1)^s \, (s-1)! \, \Big(\frac{x^2}{1-x}\Big)^s \label{Formula}
\end{eqnarray}
Here $_2F_1$ is a hypergeometric function. Proof of this identity goes as follows. We begin with a 
quadratic transformation of $_2F_1$.\cite{Grad}
\begin{eqnarray}
_2F_1(\alpha,\beta,2\beta;x) &=& \bigg(1-\frac{x}{2}\bigg)^{-\alpha} \, _2F_1\left(\frac{\alpha}{2}, 
\frac{\alpha+1}{2},\beta+\frac{1}{2};\bigg(\frac{x}{2-x}\bigg)^2\right)
\end{eqnarray}
In this formula we set $\alpha=1$ and $\beta=1-s+\epsilon$, where $\epsilon$ is an infinitesimal parameter, 
which will be sent to 0. On the right hand side, we can safely take this limit, and obtain $(2/(2-x)) \, 
_2F_1(\frac{1}{2},1,\frac{3}{2}-s;x^2/(2-x)^2)$. On the left hand side, the second and third parameter 
of the hypergeometric function become negative intergers in this limit.  By taking care of this point one obtains
\begin{eqnarray}
&& \lim_{\epsilon \rightarrow 0} \, _2F_1(1,1-s+\epsilon, 2-2s+2\epsilon;x) \nonumber \\
&=&  \frac{(s-1)!}{(2s-2)!} \, \sum_{n=1}^s \frac{(2s-n-1)!}{(s-n)!} \, x^{n-1}+\frac{1}{2} \, (-1)^{s+1} \, 
\frac{[(s-1)!]^2}{(2s-2)!} \, \frac{x^{2s-1}}{ (1-x)^s}.
\end{eqnarray}
This proves the formula (\ref{Formula}). 

When this formula is applied to the summation of two terms in (\ref{threep00}), the arguments of the two $_2F_1$'s from both terms coincide and they cancel  out. Only the second term in (\ref{Formula}) contributes and 
the three-point function is given by 
\begin{eqnarray}
<{\it O}^{(0,0)}(z_1) \, \bar{{\it O}}^{(0,0)}(z_2) \, J^{(s)}(z_3)> = \frac{g}{2\sqrt{z_{12} \, \bar{z}_{12}}} \,  \frac{4^{1-s}}{2\pi} \, (s-1)!\, \Bigg(\frac{z_{12}}{z_{13} \, z_{23}}\Bigg)^s.
\end{eqnarray}
Similarly the other three-point functions are calculated.  All the results are summarized as follows. ($\nu=0,1$)
\begin{multline}
 <{\it O}^{(0,\nu)}(z_1) \, \bar{{\it O}}^{(\nu,0)}(z_2) \, J^{(s)}(z_3)>  \\
= \frac{4^{1-s}}{2\pi} \, (s-1)!\, \Bigg(\frac{z_{12}}{z_{13} \, z_{23}}\Bigg)^s \,  
<{\it O}^{(0,\nu)}(z_1) \, \bar{{\it O}}^{(\nu,0)}(z_2)>, 
\end{multline}
\begin{multline}
<{\it O}^{(1,\nu)}(z_1) \, \bar{{\it O}}^{(\nu,1)}(z_2) \, J^{(s)}(z_3)> \\ 
=  \frac{4^{1-s}}{2\pi} \, (s-1)!\, (2s-1)\,  \Bigg(\frac{z_{12}}{z_{13} \, z_{23}}\Bigg)^s \, 
<{\it O}^{(1,\nu)}(z_1) \, \bar{{\it O}}^{(\nu,1)}(z_2)>.
\end{multline}
Compared to eq (4.51) of \cite{AKP2} with $\lambda=\frac{1}{2}$, our spin-s current $J^{(s)}$ is related to their $J^{(s)\text{AKP}}$ as $J^{(s) }=(-1)^{s+1} \, J^{(s) \text{AKP}}$. Then the results for ${\it O}^{(0,0)}$ 
and ${\it O}^{(1,1)}$ agree. This result also shows that ${\it O}^{(0,1)}$, ${\it O}^{(1,0)}$ and their partners $\bar{{\it O}}$  are primaries.

\section{Discussion}
\hspace{5mm}
In this paper we studied a model of matter fields coupled to 3d higher-spin gravity, where matter fields are a real 0-form $C$ and a two-form $B$, and found that the solutions to the classical equations of motion of our theory also satisfy the linearized equations of motion of 3d Vasiliev gravity. A local boundary action which yields two-point correlation functions in the boundary CFT is found, and a method for calculating three-point functions for two primary operators and a spin-s current  within the on-shell action method is presented. By using this method in $HS[\frac{1}{2}] \times HS[\frac{1}{2}]$ gravity, solutions to the equations of motion in AdS$_3$ bcakground are found and the two-point functions of the primary operators dual to the matter fields are obtained.  They agree with the results of linearized 3d Vasiliev gravity. The two-point functions of the fermion operators are also obtained. 

The interesting fact that the on-shell boundary term $\text{tr} \,  B \, C$ does not depend on the coordinates is a consequence of the equations of motion for matter fields, {\it i.e.}, covariantly-constancy conditions. As a result, our prescription for holography shows a novel feature: it is not necessary to take the near boundary limit. The holographic screen is not at infinity. Data of the bulk is stored in the internal space sitting at a single point in the bulk. Due to the same reason, correlation functions of all the operators with two scaling dimensions $\Delta_{\pm}$ in `boundary CFT' are obtained without doubling the real matter fields $C$ and $B$.

It will be interesting to extend the present work also to the case of backgrounds other than AdS$_3$ and BTZ black hole, such as higher-spin black hole geometry, and $HS[\lambda] \times HS[\lambda]$ gravity with arbitrary $\lambda$ within 
$0 \leq \lambda \leq 1$.  
Finally, it is a challenging problem to introduce interactions among matter fields in an invariant way under higher-spin gauge transformations.

\setcounter{section}{0}
\renewcommand{\thesection}{\Alph{section}}

\section{Notations for sl(3,R) algebra}
\hspace{5mm}
In this appendix notations related to $sl(3,R)$ algebra are summarized. 

Let the generators $L_i \ (i=-1,0,1)$, $W_n \ (n=-2, \ldots 2)$ satisfy 
an $sl(3, R)$ algebra.
\begin{eqnarray}
\ [L_i,L_j] &=& (i-j) \, L_{i+j}, \qquad 
\ [L_i,W_n] = (2i-n) \, W_{i+n}, \nonumber \\
\ [W_m,W_n] &=& -\frac{1}{3} (m-n) \, (2m^2+2n^2-mn-8) \, L_{m+n}
\end{eqnarray}
We use the following representation of SL(3,R).
\begin{eqnarray}
L_1 &=& \left(\begin{array}{ccc}
        0  & 0& 0 \\
        1 & 0 & 0 \\
        0 & 1 & 0  \end{array}\right), \qquad 
L_0  = \left(\begin{array}{ccc}
        1  & 0& 0 \\
        0 & 0 & 0 \\
        0 & 0 & -1 \end{array} \right), \qquad 
L_{-1}  = \left(\begin{array}{ccc}
        0  & -2& 0 \\
        0 & 0 & -2 \\
        0 & 0 & 0  \end{array}\right), \nonumber \\
W_2 &=& \left(\begin{array}{ccc}
        0  & 0& 0 \\
        0 & 0 & 0 \\
        2 & 0 & 0  \end{array}\right), \qquad 
W_1  = \left(\begin{array}{ccc}
        0  & 0& 0 \\
        1 & 0 & 0 \\
        0 & -1 & 0 \end{array} \right), \qquad 
W_0  = \frac{2}{3} \, \left(\begin{array}{ccc}
        1  & 0& 0 \\
        0 & -2 & 0 \\
        0 & 0 & 1  \end{array}\right), \nonumber \\
W_{-1} &=& \left(\begin{array}{ccc}
        0  & -2 & 0 \\
        0 & 0 & 2 \\
        0 & 0 & 0 \end{array} \right), \qquad 
W_{-2}  = \left(\begin{array}{ccc}
        0  & 0& 8 \\
        0 & 0 & 0 \\
        0 & 0 & 0  \end{array}\right)
\label{generators}
\end{eqnarray}
Non-vanishing norms of these matrices are given by 
\begin{eqnarray}
&&\mbox{tr} \ (L_0)^2=2, \quad \mbox{tr} \ (L_{-1}L_1)=-4, \nonumber \\
&& \mbox{tr} \ 
 (W_0)^2=\frac{8}{3}, \quad \mbox{tr} \ (W_1W_{-1})=-4, \quad \mbox{tr} \
 (W_2W_{-2})=16.
\end{eqnarray}

\section{Identities involving Moyal products}
\hspace{5mm}
In addition to the formulae in appendix A of \cite{KP2}, the following results for Moyal products are useful for the calculation in this paper.
\begin{eqnarray}
F(y_1) * e^{iy_1y_2} &=& 0, \nonumber \\ 
F(y_2) * e^{iy_1y_2} &=& F(2y_2) \, e^{iy_1y_2} , \nonumber \\
e^{iy_1y_2} * F(y_1) &=& F(2y_1) \, e^{iy_1y_2} , \nonumber \\
e^{iy_1y_2} * F(y_2) &=& 0, 
\end{eqnarray}
\begin{multline}
 e^{\frac{1}{2} \, y^T \, M \, y +\xi^T \, y} * e^{\frac{1}{2} \, y^T \, N \, y+\eta^T \, y}  \\
= \frac{1}{\sqrt{|L|}} \, \exp \Big\{ \frac{1}{2} \, y^T \, S \, y+\frac{1}{2} \, \xi^T \, (1+2X^T N-X^T \sigma_2-\sigma_2NXM) \, y \\
 +\frac{1}{2} \, \eta^T \, (1+2XM+X \sigma_2+\sigma_2MX^TN) \, y +\eta^T \, X \, \xi  \\
 -\frac{1}{2} \, \xi^T \, \sigma_2 NX \, \xi+\frac{1}{2} \, \eta^T \, \sigma_2 MX^T \, \eta \Big\}, \label{B2}
\end{multline}
where $M$ and$N$ are symmetric matrices, and $|L|= \det M \det N+\text{tr} (M\sigma_2 N \sigma_2)+1$, $X=(\sigma_2+M\sigma_2 N)^{-1}$,  $S=\sigma_2 X M+M X^T N+N X M-\sigma_2 X^T N$. Finally, 
\begin{multline}
e^{-\frac{i}{4} \, (z_1-z_2) \, y_1^2} * e^{-iy_1 \, y_2} * e^{\frac{i}{4} \, (\bar{z}_1-\bar{z}_2) \, y_2^2} * e^{iy_1 \, y_2}  \\
=   \frac{1}{2} \, (z_1-z_2)^{-\frac{1}{2}} \, (\bar{z}_1-\bar{z}_2)^{-\frac{1}{2}} \, \exp \Big(iy_1 \, y_2+\frac{i}{z_1-z_2} \, y^2_2\Big), \label{prodmp}
\end{multline}

\section{Scalar Two-point Functions on BTZ}
\hspace{5mm}
The gauge fields (\ref{asmAdS3}) for BTZ black hole are given by \cite{KP2} 
\begin{eqnarray}
a &=& (V^2_1+\frac{1}{4\tau^2} \, V^2_{-1}) \, dz, \\
\bar{a} &=& (V^2_{-1}+\frac{1}{4\bar{\tau}^2} \, V^2_{1}) \, d\bar{z}.
\end{eqnarray}
Here  $\tau$ and $\bar{\tau}$ are modular parameters of the Eucllidean boundary torus. By using  (\ref{Csol}) and (\ref{Bsol}), solutions for  $C$ and $B_{\phi\rho}$ are given by 
\begin{eqnarray}
C^{(\nu_1,\nu_2)} &=& b^{-1} * e_*^{-(z-z_1) \, (V^2_1+\frac{1}{4\tau^2} \, V^2_{-1}) } * f_{\nu_1,\nu_2}(y) * e_*^{(\bar{z}-\bar{z}_1) \, (V^2_{-1}+\frac{1}{4\bar{\tau}^2} \, V^2_{1}) } * b^{-1}, \\
B_{\phi\rho}^{(\nu_2,\nu_1)} &=& b * e_*^{-(\bar{z}-\bar{z}_2) \, (V^2_{-1}+\frac{1}{4\bar{\tau}^2} \, V^2_{1}) } * \tilde{f}_{\nu_2,\nu_1}(y) * e_*^{(z-z_2) \, (V^2_{1}+\frac{1}{4\tau^2} \, V^2_{-1}) } * b.
\end{eqnarray}
Here $b(\rho)=e_*^{\rho \, V^2_0}$ and $\nu_1, \nu_2=0,1$. 
By substituting these solutions into the boundary action (\ref{newS}), two-point functions are obtained.
\begin{eqnarray}
\langle \Psi| {\it O}^{(0,0)}(z_1,\bar{z}_1) \, \tilde{{\it O}}^{(0,0)}(z_2,\bar{z}_2) |\Psi \rangle &=& \frac{1}{4} \, g \, \Bigg(\frac{1}{ \tau \, \bar{\tau} \,  \sin \frac{z_{12}}{2\tau} \, \sin \frac{\bar{z}_{12}}{2\bar{\tau}}    }   \Bigg)^{\frac{1}{2}}, \label{thermal1}\\
\langle \Psi| {\it O}^{(1,1)}(z_1,\bar{z}_1) \, \tilde{{\it O}}^{(1,1)}(z_2,\bar{z}_2) |\Psi \rangle &=& \frac{1}{4} \, g \, \Bigg(\frac{1}{\tau \, \bar{\tau} \,   \sin \frac{z_{12}}{2\tau} \, \sin \frac{\bar{z}_{12}}{2\bar{\tau}}    }   \Bigg)^{\frac{3}{2}}, \\
\langle \Psi| {\it O}^{(0,0)}(z_1,\bar{z}_1) \, \tilde{{\it O}}^{(1,1)}(z_2,\bar{z}_2) |\Psi \rangle &=& \langle \Psi| {\it O}^{(1,1)}(z_1,\bar{z}_1) \, \tilde{{\it O}}^{(0,0)}(z_2,\bar{z}_2) |\Psi \rangle =0 \nonumber \\
&& \label{thermal3}
\end{eqnarray}
Here $z_{12}=z_1-z_2$, etc, and $|\Psi \rangle$ is an entangled state in a tensor product of two CFTs. Both operators in these two-point functions live in the same Hilbert spaces ${\cal H}_R$ or  ${\cal H}_L$:  
$\langle \Psi| {\it O}_{R,L}^{(\nu_1,\nu_2)}(z_1,\bar{z}_1) \, \tilde{{\it O}}_{R,L}^{(\nu_3,\nu_4)}(z_2,\bar{z}_2) |\Psi \rangle$. Two-point functions of a form $\langle \Psi| {\it O}_L^{(\nu_1,\nu_2)}(z_1,\bar{z}_1) \, \tilde{{\it O}}_R^{(\nu_3,\nu_4)}(z_2,\bar{z}_2) |\Psi \rangle$, {\em i.e.}, mixed correlators,  are obtained after a half shift, $z \rightarrow \pi \tau$ and $\bar{z} \rightarrow \pi \bar{\tau}$, in (\ref{thermal1})-(\ref{thermal3}). These results agree with those in \cite{KP2}. Hence our prescription with the boundary term (\ref{newS}) also works for matter fields on BTZ black hole. Furthermore, two-point functions of operators with different scaling dimensions (\ref{thermal3}) vanish.

\end{document}